\documentclass[fleqn,usenatbib]{mnras}
\usepackage{newtxtext,newtxmath}
\usepackage[T1]{fontenc}
\usepackage{ae,aecompl}
\usepackage[usenames, dvipsnames]{color}
\usepackage{graphicx}	
\usepackage{amsmath}	
\usepackage{amssymb}	

\newcommand{\cmps}{\ensuremath{\textrm{cm}~\textrm{s}^{-1}}}
\newcommand{\mps}{\ensuremath{\textrm{m}~\textrm{s}^{-1}}}
\newcommand{\kmps}{\ensuremath{\textrm{km}~\textrm{s}^{-1}}}
\renewcommand{\th}{\textsuperscript{th}} 

\newcommand{\mystar}{HD\;127334}

\newcommand{\lrhk}{\ensuremath{\log R'_{\rm HK}}}

\newcommand{\rev}[1]{#1}

\title[Robust, template-free \& precise RV extraction]{A robust, template-free approach to precise radial velocity extraction}
\author[V.~M.~Rajpaul et al.]{
V.~M.~Rajpaul,$^{1}$\thanks{E-mail: vr325@cam.ac.uk}
S.~Aigrain$^{2}$, and
L.~A.~Buchhave$^{3}$
\\
$^{1}$University of Cambridge, Astrophysics Group, Cavendish Laboratory, J. J. Thomson Avenue, Cambridge CB3 0HE, UK\\
$^{2}$Sub-department of Astrophysics, Department of Physics, University of Oxford, Oxford OX1 3RH, UK\\
$^{3}$DTU Space, National Space Institute, Technical University of Denmark, Elektrovej 328, DK-2800 Kgs.\ Lyngby, Denmark\\
}

\date{Accepted ---. Received ---; in original form ---}

\pubyear{2019}

\begin{document}
\label{firstpage}
\pagerange{\pageref{firstpage}--\pageref{lastpage}}
\maketitle

\begin{abstract}
Doppler spectroscopy is a powerful tool for discovering and characterizing exoplanets. For decades, the standard approach to extracting radial velocities (RVs) has been to cross-correlate observed spectra with a weighted template mask. While still widely used, this approach is known to suffer numerous drawbacks, and so in recent years increasing attention has been paid to developing new and improved ways of extracting RVs. In this proof-of-concept paper we present a simple yet powerful approach to RV extraction. We use Gaussian processes to model and align all pairs of spectra with each other; without constructing a template, we combine pairwise RV shifts to produce accurate differential stellar RVs. Doing this on a highly-localized basis enables a data-driven approach to identifying and mitigating spectral contamination, even without the input of any prior astrophysical knowledge. We show that a crude implementation of this method applied to an inactive standard star yields RVs with comparable precision to and significantly lower rms variation than RVs from industry-standard pipelines. Though amenable to numerous improvements, even in its basic form presented here our method could facilitate the study of smaller planets around a wider variety of stars than has previously been possible.
\end{abstract}




\section{Introduction}\label{sec:intro}

The discovery of exoplanets has heralded the start of one of history's greatest scientific revolutions, and exoplanetary science has rapidly acquired a unique positioned to address profound questions about the origins of life, and about our place and future in the cosmos.

Since the discovery of the first exoplanet over two decades ago \citep{mayor1995}, Doppler spectroscopy -- a.k.a.\ the radial velocity (RV) method -- has been one of the bedrocks of exoplanetary science. As of 2019, it has been responsible for more exoplanet discoveries than all other techniques apart from transit photometry; and together, the RV and transit methods have been responsible for the discovery of \rev{around} $95\%$ of all confirmed exoplanets.\footnote{Based on counts from the NASA Exoplanet Archive, available online at \url{exoplanetarchive.ipac.caltech.edu}.} Apart from being an important tool for exoplanet discovery in its own right, the RV method has also been indispensable for confirming and characterizing planetary candidates discovered via other techniques \citep[e.g.][]{konacki03}: of particular importance is the RV method's ability to constrain a planet's mass, and thus provide information about its likely composition, formation history, atmosphere scale height, and more. 


Thanks to a number of technical advances, the precision of RV surveys has been steadily improving. Whereas the RV spectrographs of fifty years ago yielded RV measurements with nominal errors in excess of $1$~\kmps\ per measurement, absorption-cell spectrographs have in recent years demonstrated precisions of order $1$~\mps\ \citep{butler2017}, while the newest generation of highly-stabilized spectrographs today boast sub-\mps\ precisions, and aim to achieve $10$~\cmps\ precisions \citep{Pepe2010,probst2014,HARPS3,jurgenson2016,schwab2016}. Although absorption-cell spectrographs \citep{cwalker79,marcybutler92,butler1996} were responsible for most of the first several dozen exoplanet discoveries, more recently ultra-stabilized spectrographs \citep[i.e.\ using the so-called `simultaneous reference' technique:][]{Bara96,probst2014} have yielded the highest precisions. The most ambitious plans for next-generation RV instruments call for stability at the $1$~\cmps\ level \citep{pasquini2008,fischer2016eprv}.

Despite enormous advances in instrumentation, a number of significant obstacles continue to impede the discovery of low-mass planets, especially below the $\sim1$~\mps\ level. Moreover, 10~\cmps\ precisions have not yet been demonstrated.

Perhaps the most vexatious obstacle to detecting true Earth-analogue planets using Doppler spectroscopy (or indeed transit photometry) is variability intrinsic to stars themselves. These stellar nuisance signals, due e.g.\ to surface magnetic activity such as rotating spots and plages, may have covariance structure similar to but amplitudes orders of magnitudes larger than the signals expected from true Earth-analogues \citep{dumusque2012}. There has thus been considerable effort devoted, in recent years, to developing ways to disentangle stellar activity signals from planetary ones in RVs \citep[][]{boi+09,lan+10,aigrain2012,tuomi2014,haywood2014b,robertson2014,Rajpaul2015,jones2017}. With one or two exceptions \citep[e.g.][]{davis17}, almost all such efforts have been based on \emph{post hoc} attempts to constraining the stellar activity component of RVs. In other words, RVs derived via a standard pipeline are taken as a starting point, and supplementary information is used after the fact (e.g.\ ancillary photometry, or spectroscopic diagnostics that should be sensitive to activity but not planets) to try to figure out which RV variability may be due to stellar activity, and which to planets. Despite modest progress in this direction, there remains significant scope and indeed compelling need for improved approaches to mitigating stellar activity's impact on RV exoplanet detection and characterization.

Telluric absorption and emission lines pose another non-trivial obstacle to extreme-precision RVs. A common approach to overcoming this obstacle is simply to discard those portions of spectra known \emph{a priori} to suffer from significant atmospheric contamination. This, however, has the effect of throwing away potentially-valuable information that could be used to constrain genuine stellar RVs. Moreover, it is not straightforward to mask off the forest of so-called `micro-telluric' features which, though almost imperceptible, can nevertheless lead to RV contamination at the meter per second level, depending on factors including atmospheric water vapour content and air mass \citep{cunha14}. Correction using standard stars \citep{vacca03} or theoretical atmospheric transmission spectra models \citep{hrudkova05,bailey07,cotton14} is possible albeit non-trivial, and the results are imperfect. A recent and sophisticated effort to constrain telluric spectra directly from a large number of observed spectra has shown significant promise \citep[][see also Section~\ref{sec:comparison} of this paper]{wobble19}.

Against the backdrop of improving instrumentation and understanding of spectral contamination, it is interesting to note that \rev{for several decades, the standard approach to extracting RVs from spectra taken with a stabilized spectrograph remained essentially unchanged}: \emph{viz.}, cross correlating observed spectra with a template \citep{griffin67,simkin74,baranne79,tonry79,bouchy01}. This is typically either a synthetic template based on model stellar atmospheres, knowledge of atomic line locations, and other such considerations, or a high signal-to-noise ratio (SNR) spectrum derived from real observations \citep[e.g.][]{Nord94,Bara96,balona02}, in either case with various numerical weights and/or masking applied to different parts of the template \citep{pepe02}. \rev{While increasing attention has been paid in recent years to developing better approaches to RV extraction \citep[e.g.][]{anglada12,zech2018,dumusque18,wobble19}, the approach of cross-correlating observed spectra with a masked, weighted template (often called a \emph{delta function} template when a binary mask is used) retains wide currency. For example, this approach is employed in the primary data reduction pipelines of HARPS \citep{rupprecht04} and HARPS-N \citep{cosentino12}, as well as in the pipelines of newer instruments such as ESPRESSO \citep{espresso18} and EXPRESS \citep[e.g.][]{blackman19}.}

Though tried and trusted, this standard approach suffers several drawbacks. First, a pre-defined template will never be a perfect match to any observed star's spectrum, be it because the template is derived from a different star, or because the line locations and depths in a synthetic template are known with only finite precision. Masking also inevitably throws away some RV information contained outside of the lines chosen for the template. Second, the RV extraction process does not, by itself, suggest strategies for mitigating stellar activity variability and telluric contamination. Third, errors in RVs extracted via template cross-correlation tend to be estimated via ad hoc stratagems \citep[see e.g.][]{tonry79} that do not account for template imperfections, spectral contamination, etc. Fourth, \rev{when using a pre-defined template}, acquiring more spectra does not improve the accuracy or precision of existing RVs, despite additional spectra containing potentially-useful new information (each observed spectrum could be used to reduce noise and/or improve knowledge of line locations in the template); instead, each RV is computed via a once-off cross-correlation, independent of all other observations. \rev{Alternatively, as we discuss in Section~\ref{sec:pairwise-xcorr1}, there are numerous issues inherent in using observed spectra to construct a `master' template.} Finally, one could argue that this whole process is remarkably complex: depending on the particular implementation, RV extraction might ultimately involve everything from atomic physics and stellar atmosphere modelling (in the context of template construction) to terrestrial meteorological considerations (in the context of telluric masking or modelling) to an assortment of numerical schemes for resampling, interpolation, error propagation, etc. 

In practice, one \rev{does not actually concern oneself with most of these complexities -- typically, one simply passes observed spectra to off-the-shelf code}, and after `turning the crank,' ends up with a set of RVs and error estimates. But building an RV extraction pipeline from the ground up, or diagnosing shortcomings of an existing approach, would require \rev{vastly more} work and insight. 

In this paper, we add to the growing number of recent efforts to develop new and improved ways of extracting RVs, and propose a conceptually simple, Gaussian process-based approach to extracting RVs from ultra-stabilized spectra. Our approach uses nothing but non-parametric Gaussian process (GP) models of observed spectra for RV extraction; it requires no template, nor any inputs of prior astrophysical or telluric knowledge. 
Despite its simplicity, we show that our technique avoids most of the issues we enumerated previously, and indeed is able to outperform a number of currently-used approaches to RV extraction.

\section{A template-free approach to radial velocity extraction}\label{sec:technique-sketch}

\subsection{The basic idea}

We present here the simple idea behind our proposed new technique; we shall afterwards discuss and motivate the technique's features. Given $N$ spectra, the essence of the technique is the following:
\begin{enumerate}
	\item assume each observed spectrum can be modelled with a GP;
	\item align all GP model spectra to infer RV shifts (plus associated error estimates) between all pairs of spectra;
	\item \rev{selectively combine} these $\mathcal{O}(N^2)$ pairwise RV shifts, accounting for estimated errors, to produce $N$ differential RVs.
\end{enumerate}

\begin{figure*}
	\begin{center}
		\includegraphics[width=\textwidth]{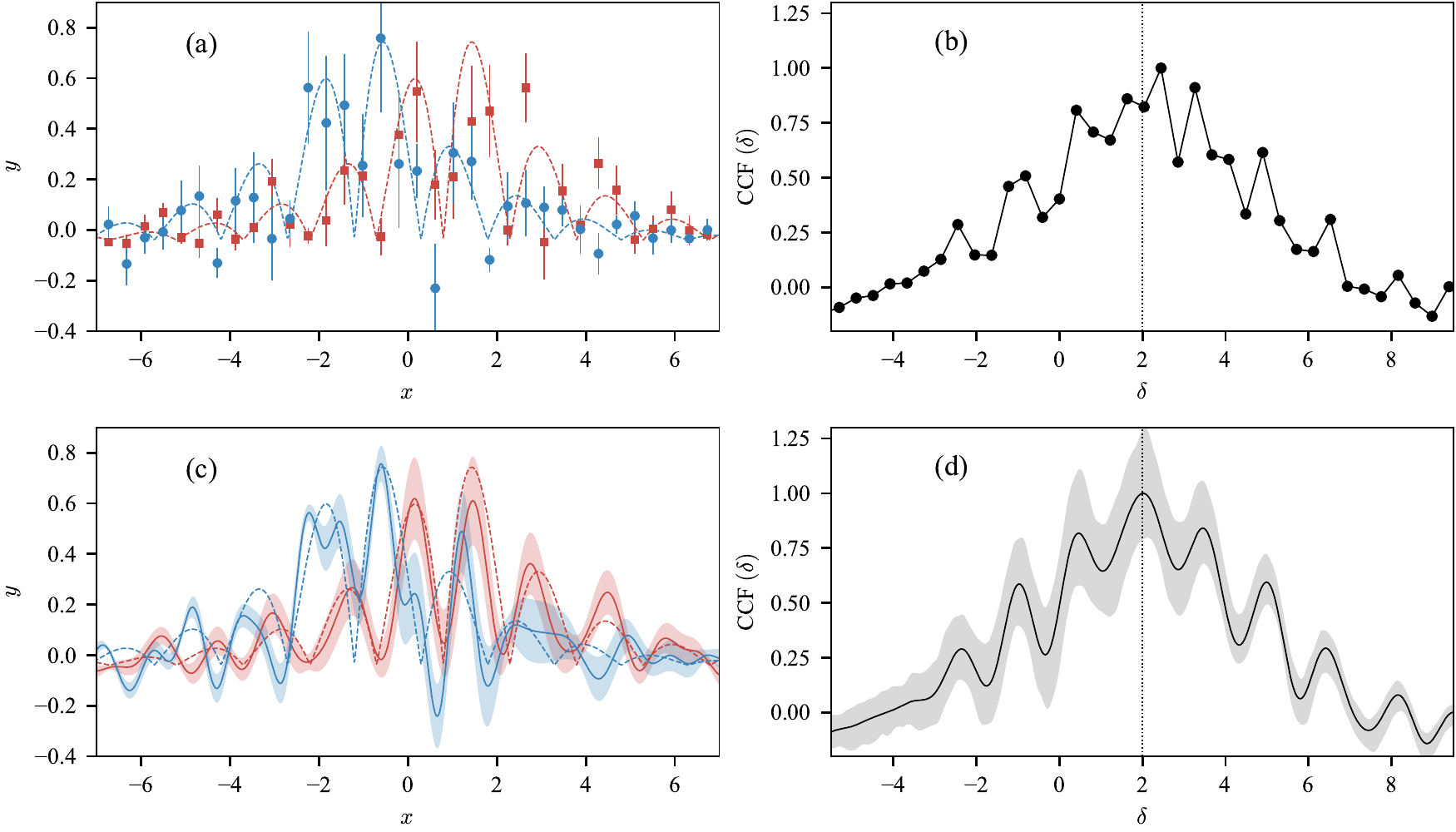}
		\caption[Caption]{Cross-correlation using discrete, noisy data vs.\ GP models for the data. Panel (a) shows two sets of noisy observations of an arbitrarily-chosen function; the dotted lines show the noise-free function (an exponentially-decaying squared sinusoid) used to generate the observations, which are in turn denoted by red and blue points. A lag of $\delta=2$ was used. Panel (b) shows the discrete, normalized cross-correlation function (CCF) computed directly from the observations; the black dotted line indicates $\delta=2$. Panel (c) shows GP models with Gaussian kernels fitted to the observations; the solid lines and shaded regions denote posterior means and $\pm\sigma$ posterior uncertainty, respectively. The GP means recreate the original generating functions quite well. Panel (d) shows the normalized CCF computed from the models in panel (c). The CCF is now continuous and attended by a detailed uncertainty model, which automatically accounts for the uneven information content of the cross-correlated observations, gaps in the observations, etc.}
		\label{fig:xcorr_vs_GP_xcorr}
	\end{center}
\end{figure*}

\subsection{Motivation for using GPs}\label{sec:GP-motivation}

Why bother modelling spectra with GPs, before inferring RV shifts between them?\footnote{Readers looking for a quick, conceptual introduction to GPs are referred to, e.g., \citet[][]{roberts2013} or Section 2.6 of \citet{vinesh-thesis}.} Why not just align the `raw' spectra directly?

In practice, each observed spectral flux will be accompanied an error estimate, based at a minimum on Poisson counting statistics associated with photon noise. However, this belies the fact that except in regions of pure noise, errors will generally be correlated along the wavelength axis: knowledge of a flux at one point should \emph{reduce} our uncertainty in nearby flux values. For example, if a spectral line is traced out over (say) ten detector pixels, these ten fluxes combined with some knowledge of the regular rather than completely random structure of the line should allow us to interpolate the fluxes more accurately compared to the case where the fluxes were treated as arising purely from white noise. Of course, this behaviour would be localized: knowledge of fluxes in a given line wouldn't tell us anything useful about distant lines; i.e., covariance length scales would be short.

Provided the covariance length scale is finite, though, we might as well work with a covariance matrix quantifying how knowledge of fluxes at a given set of points reduces uncertainty about nearby fluxes, instead of treating flux errors as being independent. In other words, we might as well work with GPs, which in one sense are nothing but a generalisation of white Gaussian noise models. Using a GP with suitable covariance function permits a more realistic (less pessimistic) error treatment than when assuming independent error estimates on individual fluxes, which should ultimately lead to refined RV estimates. Given that RVs of $1$~\mps\ translate into shifts across about 1/1000\th\ of a detector pixel on an instrument like HARPS or HARPS-N, we shouldn't be surprised if even a slight improvement to something as mundane as spectral interpolation could translate into improved RV estimates.

Using GPs, we need not assume any detailed knowledge of or parametric form for the spectrum/function we wish to learn, and we certainly do not need to input any atomic or stellar astrophysics. It suffices for the spectrum to have some non-random structure (absorption lines, continuum, etc.), and thus non-trivial covariance structure. Provided we can \emph{parametrize} this covariance and thus formulate a prior over functions, \rev{we can then use the elegant machinery of GP regression to infer analytically a function to describe the observations, along with detailed error bars on that function, encoded in the GP posterior mean and covariance, respectively.} 

\rev{Once working within a GP regression framework, it becomes straightforward to evaluate and manipulate model spectra and associated uncertainties at arbitrary wavelengths, rather than just the discrete set of wavelengths defined by detector pixels, and even to apply a variety of operators analytically. For example, the joint GP likelihood for the two spectra as a function of relative RV shift can be evaluated analytically and maximised directly, without having to worry separately about resampling spectra to a common wavelength grid, propagating uncertainties, etc. Alternatively, GPs might be cross-correlated to produce a continuous CCF attended by a detailed uncertainty model; see Fig.~\ref{fig:xcorr_vs_GP_xcorr} for an example. These two (related) approaches to RV extraction are considered in Section~\ref{sec:RV_extraction}.}

Of course, there are countless other ways one could model spectra non-parametrically. However, GPs offer a principled way to do this modelling, and \rev{as noted above} also conveniently take care of most of the things one would anyway need to do when inferring RVs, \rev{whether via cross-correlation or a maximum-likelihood (ML) approach}. Moreover, as we shall discuss in Section~\ref{sec:future}, the basic GP framework we present here will prove amenable to various sophisticated extensions, including explicit modelling of activity and telluric variability via modelling of temporal covariances.

\subsection{\rev{Motivation for computing pairwise RVs}}\label{sec:pairwise-xcorr2}

\rev{Before turning to the practical matter of \emph{how} to aggregate an ensemble of pairwise RV shifts to produce differential RVs, it is useful first to ask the question: \emph{why} compute pairwise RVs at all? Why not take a more traditional approach and align each spectrum with a template formed by co-adding all spectra? Or why not align all spectra simultaneously, instead of doing it on a pair-by-pair basis?}


\rev{First, there are theoretical reasons to prefer the pairwise-RV approach. Even if portions of two different spectra contain stellar activity and telluric contamination, it will still be possible to compute a single, well-defined relative RV shift between the two spectra, at least when considering the uncontaminated portions of the spectra.}

\begin{figure}
	\begin{center}
		\includegraphics[width=\columnwidth]{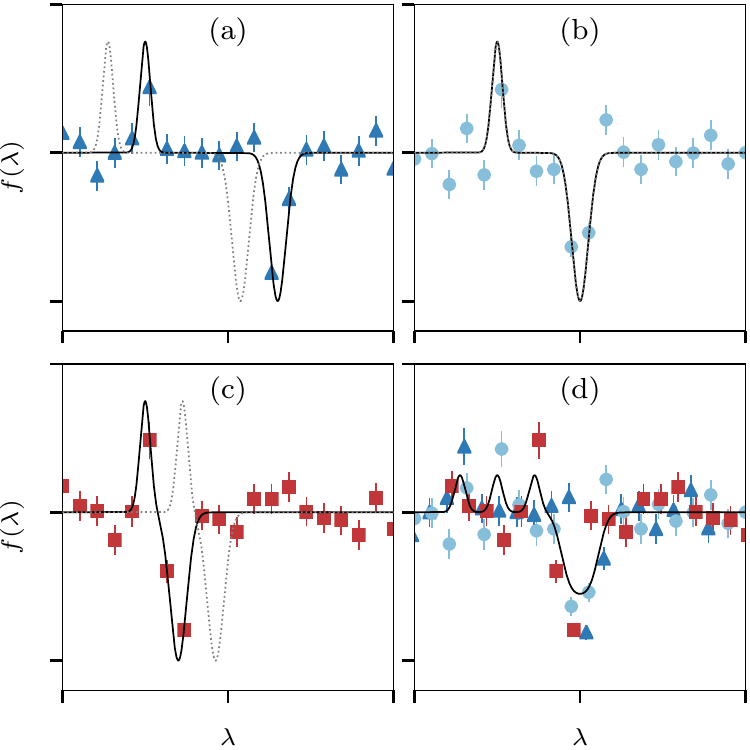}
		\caption[Caption]{\rev{Schematic of a spectrum $f(\lambda)$ comprising  stellar absorption line, telluric emission feature, and photon noise. The stellar line has positive, zero and negative RV in panels (a) to (c), and for the purposes of a simple illustration, Earth's barycentric velocity is assumed to have the same sign but three times the amplitude of the stellar velocity. The solid and dotted lines indicate the noise-free spectrum before and after barycentric velocity correction. Co-adding the barycentric velocity-corrected observations yields the average `template' in panel (d). Per the discussion in Section~\ref{sec:pairwise-xcorr1}, this template does not resemble any of the original spectra up to an RV shift: telluric contamination is spread out, while the stellar line is made shallower and broader by adding observations with non-zero relative stellar RVs.}}
		\label{fig:telluric-shift}
	\end{center}
\end{figure}

\rev{On the other hand, imagine a large number $N$ of stellar spectra that have non-zero relative RVs, due e.g.\ to the presence of orbiting planets or a stellar companion, but are otherwise identical. A template formed by co-adding all spectra will smear out sharp spectral features in proportion to the relative stellar RVs, possibly blending nearby lines that were not blended in individual spectra, etc. If, additionally, stellar activity contamination is present, lines in the co-added template will exhibit broadening, skewness, etc.\ from co-adding time-variable activity distortions \citep{Boi11} across a large number of spectra. Similarly, if telluric contamination is present, the co-added template will contain telluric features smeared out by up to $\pm30$~\kmps\ after barycentric correction for Earth's annual orbital motion, and by up to $\pm460$~\mps\ after barycentric correction for Earth's daily rotation \citep{wright2014}, irrespective of any genuine stellar Doppler shifts that may be present.\footnote{\rev{Earth's motion has an unexpectedly useful consequence for Doppler spectroscopy: over time, a single detector will probe an ever-denser set of stellar photon \emph{emission} wavelengths, despite the detector's finite number of pixels with well-defined central wavelengths. This can be exploited to build up a template stellar spectrum with resolution superior to that of any individual spectrum  \citep{park2003,grundhofer2017}. Our template-free technique does this implicitly through RV estimation from pairs of spectra with different emission wavelength sampling.}} All of which means that the template will never be equivalent up to a velocity shift to \emph{any} of the individual spectra, translating into sub-optimal RV extraction. See Fig.~\ref{fig:telluric-shift}.}

\rev{A more subtle problem is that because the noise in any individual spectrum will be present albeit diluted in the co-added template, there will always be a small bias towards zero RV due to noise in every spectrum matching with itself in the template \citep{kerkwijk95}. This bias decreases with increasing $N$, though could be significant even for large $N$ in the case of non-Gaussian outliers.}

\rev{In principle there are ways to address these issues while using an observed template, though they are usually not straightforward. For example, even without activity or telluric contamination, there would inevitably be a `chicken and egg' problem to solve: to build an optimal template would require un-shifting and stacking observed spectra; but knowing by how much to un-shift a given spectrum would itself require comparison with a template. Hence an iterative \citep[e.g.][]{cabasson06,rezk12} scheme for refining estimated shifts would needed; even if these schemes converge quickly, there is generally no guarantee that they will converge to the globally optimal solution. Separate strategies would be needed for modelling or masking off activity and/or telluric contamination. Bias related to self-matching of noise could be avoided by working with a separate template for each spectrum (excluding that spectrum itself from the template), although this would immediately increase the complexity and computational burden of the extraction by an order of magnitude, since now $N$ different templates would need to be constructed iteratively. Or one could simply ignore these issues, though to do so would be contrary to the pursuit of extreme-precision RVs.}

\rev{On the other hand, assume that we are working with only two spectra, and wish to compute a relative RV shift for different `chunks' (e.g.\ \'echelle orders or smaller subsets) of the spectra. Now, the local RV shifts we infer will fall into one of two categories: (i) \emph{bona fide} stellar Doppler shift estimates, from regions in the pair of spectra that are equivalent up to a Doppler shift; and (ii) `bad' RV estimates from regions in the pair of spectra that differ on account of tellurics, activity contamination, or instrumental systematics. Provided the latter contamination is at least somewhat localized and does not corrupt the majority of the spectrum, we should be able to identify RV estimates of type (ii) as being drawn from a very different distribution to those of type (i), and could exclude the associated wavelengths from the final RV estimation.}

\rev{If instead of constructing a template from observations we performed RV extraction by comparing observed spectra with a template derived from a \emph{different} star, or with a binary-mask template based on theoretical positions and widths of stellar absorption lines at zero velocity}, there will be additional sources of error, e.g.\  from a mismatch between stellar types (in general, the template will not match exactly the spectral type, metallicity, activity level, rotation speed, etc.\ of the target star), or from uncertainty in depths and positions of lines in the delta function template (typically hundreds of \mps\ per line), which introduces an artificial velocity offset to each order based on the ensemble of lines used in the template. The latter problem is exacerbated by the varying SNR across orders (lower at order edges than centres, due to a spectrograph's blaze function): when lines move across an order due e.g.\ to Earth's barycentric motion, the effective weight of the lines in the template changes, leading to further spurious RVs.

\rev{The numerous problems inherent in template-based RV extraction aside, there are also more practical reasons to favour pairwise RV comparison. Simultaneous alignment of many spectra, alongside telluric and activity masking or modelling, could become computationally intractable as $N$ becomes large, not least because of the enormous dimensionality of the space of free parameters. Additionally, the very large number of observations to be modelled simultaneously would certainly be an obstacle to GP-based approaches to RV estimation. By contrast, pairwise RV estimation entails cheap, repeated computation that parallelises trivially, with only a few parameters to be optimized for any pair of spectra.}



\subsection{Averaging over pairwise RVs}\label{sec:pairwise-xcorr1}

\begin{figure}
	\begin{center}
		\includegraphics[width=\columnwidth]{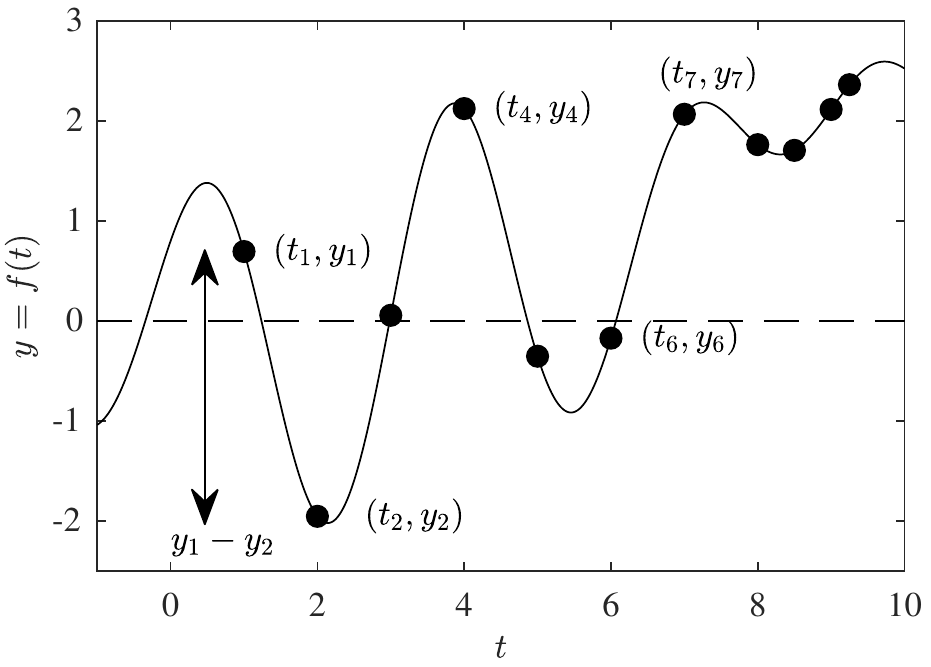}
		\caption[Arbitrary function represented as a sum of differences]{An arbitrary function, $y(t)$, with values measured relative to a fixed origin (dashed line) or via a sum of pairwise differences (see text on pg.~\pageref{pg:sum-diff}).}
		\label{fig:fxn-averaging}
	\end{center}
\end{figure}


\rev{After we compute an ensemble of pairwise RV shifts, how do we aggregate these into a `master' RV estimate for a given epoch, in the absence of reference (template) spectrum with known RV?}

As an aside, let us consider how any function can be expressed as a sum of pairwise differences.\label{pg:sum-diff} Suppose we have a function $y(t)$, with its values measured at $N$ points, so that we have a data set $\{ ({t_i},{y_i},{\sigma _i})\} _{i = 1}^N$. Here, $y_i$ is the value of the function relative to some origin (zero), and $\sigma_i$ is the standard deviation of the Gaussian error distribution for each $y_i$.

Suppose now that rather than making a single absolute measurement of $y_i$ relative to a fixed origin, we are instead able to measure the $N(N-1)/2$ pairwise differences\footnote{Recalling the well-known `handshake problem,' we only need to compute $N(N-1)/2$ pairwise differences, as $y_{ii}=0$ and $y_{ij}=-y_{ji}\;\forall\;i,j$.} between all values, so that our data would take the form $\{ ({t_{ij}},{y_{ij}},{\sigma _{ij}})\} _{i,j = 1}^N$, where $y_{ij}:=y_i-y_j$. A little algebra reveals that we can then reconstruct the individual $y_i$ values, relative to an origin defined by the average\footnote{To account for errors, we should really be weighting each $y_i$ by $\sigma_i^{-2}$ which we will indeed do later on. To simplify the discussion here, we consider for now the case of simple arithmetic averaging.} of all $y_i$ values:
\begin{equation}
\label{eq:gamma-origin}
{y_i} = \frac{1}{N}\sum\limits_{j = 1}^N {{y_i}}  = \frac{1}{N}\sum\limits_{j = 1}^N {({y_i} - {y_j}} ) + \frac{1}{N}\sum\limits_{j = 1}^N {{y_j}}  = \frac{1}{N}\sum\limits_{j = 1}^N {{y_{ij}}}  + \gamma
\end{equation}
where $\gamma:=\tfrac{1}{N}\sum\nolimits_{j = 1}^N {{y_j}} $ is our new origin. As $\gamma$ is the same for all $y_i$, if we are not interested in this offset, we may as well set $\gamma=0$. 

Why bother? In the ideal case where our origin (zero-point) is not subject to any error, it would make little sense to reconstruct $y(t)$ via pairwise differences; the individual measurements $y_i$ would, if available, constitute everything we knew about $y(t)$, and aggregating $y_{ij}$ measurements would simply reconstruct the known $y_i$ values via a pointlessly circuitous calculation. However, we could envisage two scenarios where the pairwise-differences approach would have non-trivial value. First, if the zero-point were not known at all, then we could at least use measurements of $y_{ij}$ to reconstruct $y(t)$ up to some additive constant. Second, if for some reason the measurements $y_i$ were subject to larger errors than the $y_{ij}$ measurements -- perhaps because the origin itself were poorly constrained -- then we could reconstruct $y(t)$ more accurately via $y_{ij}$ than via $y_{i}$ measurements, again up to some additive constant.

Now we could of course choose to interpret $y(t)$ as a time-varying stellar RV signal, and $y_{ij}$ as pairwise \rev{RV shifts between} spectra taken at different times. Regardless of the complexity of the RV signal $y(t)$ -- it might comprise binary stellar, exoplanetary, and other contributions -- we could reconstruct it, up to an additive constant, by averaging over the pairwise RV shifts $y_{ij}$. For the purposes of exoplanet detection or characterization, the absolute RV offset $\gamma$ would in any case be irrelevant -- so, even without any zero-RV template spectrum, we should still be able to perform precise Doppler spectroscopy. In particular, it might well be the case that $y_i$-type measurements (i.e., RVs estimated by comparing individual spectra with a fixed though imperfect template) may suffer from larger errors than $y_{ij}$-type measurements (i.e., RVs estimated by comparing pairs of real spectra from the same star), in which case we could expect the pairwise-aggregating of RV shifts to yield more accurate and precise RVs than when using an imperfect template spectrum. In fact, though we need never actually work with or compute a template, the pairwise RV shifts and uncertainties implicitly contain the information necessary to construct an ideal template derived from all the observations.


Note that provided any of the error processes over which we are averaging is stationary and has a well-defined mean, it does not matter whether the mean of the process is zero; if non-zero, the mean will simply be absorbed into the arbitrary $\gamma$ term.\footnote{Stationarity itself is not guaranteed. For example, long-term changes in a spectrograph could lead to changes in the properties of the error process. If such behaviour is suspected, it should of course be modelled, regardless of the specific technique used for extracting RVs.} 

\rev{While the error in any \emph{single} pairwise RV estimate $y_{ij}$ may well be much larger than in the case where a high-quality template were available, the standard error of the \emph{mean} RV estimates $y_{i}$ should decrease roughly in proportion with $\sqrt {1/N}$, in accordance with well-known results for the uncertainty (standard error) in the sum or mean of uncorrelated random variables \citep[e.g.][]{wasserman04}}\rev{. In other words, more spectra will mean more precise RVs, up to whatever limit is dictated by the finite information content of any individual spectrum to which wish to assign an RV shift. Of course, the errors in pairwise RV shifts will in general not be uncorrelated, especially when considering spectra closely-spaced in time, so various corrections might be made to quantify the rate at which errors should theoretically decrease, taking into account autocorrelation \citep{anderson1994,bence1995}.} We shall content ourselves with numerical experiments, presented in Section~\ref{sec:synth-data}, to investigate whether our technique, in practice, does at least approach the idealised $\sqrt {1 /N}$ improvement.

\section{Echelle spectroscopy formulation}\label{sec:echelle}

So far we have given a qualitative overview of our proposed technique, and \rev{motivated} its various features. We now flesh out the technique with enough mathematical detail to permit computational implementation, given a set of spectra taken with a stabilized \'echelle spectrograph (e.g. HARPS, HARPS-N or ESPRESSO).

\rev{Suppose we have a set of $N$ spectra, $\{ {\mathcal{S}_1}, \ldots ,{\mathcal{S}_N}\} $, where the $i$\th\ spectrum is specified by the vectors $\boldsymbol{\lambda}_i, \mathbf{f}_i, \boldsymbol{\sigma}_i \in \mathbb{R}^{M}$, containing wavelengths, fluxes, and flux error estimates, respectively. Here, $M$ is the number of pixels in a single spectrum (for HARPS-N, $69$~\'echelle~orders of $4096$~px each means $M\approx280$\;$000$~px).} We assume these wavelengths have already been transformed to take into account the relevant relativistic barycentric velocity corrections \citep{wright2014}, instrumental drift corrections, etc.

For each individual spectrum, we can try to build a model $f_i$ that aggregates information about pixel-level noise \emph{and} covariances between nearby pixels to yield a lower-noise model for resampling and cross-correlation. After all, especially in deep spectral lines, nearby pixels will \emph{not} be independent, and we could exploit this information deciding the best way to interpolate the noisy fluxes. Let us therefore place a GP prior on the function $f_i$, i.e.\ ${f_i}(\lambda )\sim\mathcal{G}\mathcal{P}({\mu _i},{k_i})$. The mean function $\mu_i(\lambda)$ will describe the continuum -- in practice, a linear or quadratic polynomial will suffice, at least within individual \'echelle orders -- while the covariance kernel will be used to build a model for the spectral absorption features. Therefore, some care is needed in the choice of covariance kernel.

\subsection{Choice of GP kernel}\label{sec:kernel}

Tests with a sample of HARPS spectra, with mean SNR at $550$~nm ranging from $50$ to $200$, revealed that a Mat\'ern kernel with $\nu=\tfrac{5}{2}$ (see Appendix~\ref{sec:app-kernel} for a definition and discussion of Mat\'ern kernels) did a very good job of fitting absorption features across all wavelengths. Setting $\nu=\tfrac{3}{2}$ led to excessive flexibility, i.e.\ obvious fitting of noise, while $\nu\ge\tfrac{7}{2}$ led to functions clearly too smooth to fit sharp absorption lines properly. Thus, setting $\nu=\tfrac{5}{2}$ in  equation~\ref{eq:matern-kernel}, and building the estimated flux errors into the kernel, our `Goldilocks' covariance function takes the form
\begin{equation}
k(\lambda ,{\lambda ^{'}}) = {h^2}\left( {1 + \frac{\tau }{\rho } + \frac{{{\tau ^2}}}{{3{\rho ^2}}}} \right)\exp \left( { - \frac{\tau }{\rho }} \right) + {\delta _{\lambda ,{\lambda ^{'}}}}{\sigma ^2}(\lambda )
\label{eq:goldilocks-kernel}
\end{equation}
where for convenience we define $\tau : = \sqrt {5{{(\lambda  - {\lambda ^{'}})}^2}}$, ${\delta _{\lambda ,{\lambda ^{'}}}}$ is the Kronecker delta, and ${k}({\lambda},{\lambda^{'}})$ is to be interpreted as the covariance between fluxes at wavelengths $\lambda$ and $\lambda^{'}$. The scale hyperparameter $h$ is related to the typical depth of spectral absorption lines, while $\rho$ is related to the line shape and typical density of lines per wavelength. \rev{To simplify notation in equation~\ref{eq:goldilocks-kernel} we have suppressed the subscript $i$ indexing the spectra; in general, the wavelengths, errors and hyper-parameters will be different for different spectra.}

As it turns out, the assumption of a GP prior enables analytic marginalizing over infinitely many unobserved function values, so that the log marginal likelihood for observations can be written down directly. If we assume our covariance kernel is controlled by some hyper-parameters $\boldsymbol \theta_i$ (i.e., $h$, $\rho$ in equation \ref{eq:goldilocks-kernel}) and our mean function by some other parameters $\boldsymbol \phi_i$ (e.g.\ polynomial coefficients), then we can write down the following straightforward expression for $\log {\mathcal{L}_i } = p({\mathcal{S}_i}| \boldsymbol \theta_i,\boldsymbol \phi_i)$:
\begin{equation}
\log {{\cal L}_i} =  - {\textstyle{1 \over 2}}\log \det {{\bf{K}}_i} - {\textstyle{1 \over 2}}{{\bf{r}}_i}^{\rm{T}}{{\bf{K}}_i}^{ - 1}{{\bf{r}}_i} - {\textstyle{M \over 2}}\log 2\pi ,
\label{eq:NLL_GP}
\end{equation}
where $\mathbf{K}_i\in\mathbb{R}^{M\times M}$ is the covariance matrix whose elements are computed by evaluating equation~\ref{eq:goldilocks-kernel} for all pairs of wavelengths in $\boldsymbol{\lambda}_i$, and $\mathbf{r}_i$ is a vector of residuals formed by subtracting the mean function evaluated at $\boldsymbol{\lambda}_i$ from the observed fluxes $\mathbf{f}_i$. Equation~\ref{eq:NLL_GP} thus allows the (hyper)parameters in $\boldsymbol \theta$ and $\boldsymbol \phi$ to be learned through an optimization routine, or indeed through posterior exploration if equation~\ref{eq:NLL_GP} is combined with suitable (hyper)priors. 

For HARPS-N FGK spectra covering the wavelength range between $383$ and $693$~nm, we found typical values of $\rho\sim0.25\pm0.05$~\AA, and for spectra normalized to have zero-mean and unit variance, $h\sim0.35\pm0.05$ (we give values for the normalized case as in general, $h$ will depend on the photon counts and variability in a given spectrum -- or part thereof, as we consider below).

Once the hyperparameters have been learned, we can compute the posterior predictive distribution, which will itself be a GP: ${g_i}(\lambda):={f_i}(\lambda )|{\mathcal{S}_i}\sim\mathcal{G}\mathcal{P}(\mu _i^*,K_i^*)$, where $\mu _i^*$ and $k_i^*$ may be computed using the prescriptions in equations 3.8 and 3.9 from \citet{roberts2013}. For brevity and to avoid defining notation we shall not again use in this paper, we do not reproduce these canonical prescriptions here; suffice it to say, though, that simple linear algebra yields $\mu _i^*$ and the covariance matrix ${{\mathbf{K}_i}^*}$ defined by $k_i^*$. Whereas the `raw' spectrum $\mathcal{S}_i$ was defined by a discrete set of noisy fluxes, $g_i$ may be evaluated at arbitrary wavelengths that interpolate $\boldsymbol{\lambda}_{i}$, and such evaluations will be accompanied by a principled flux uncertainty model.


\subsection{Comparing spectra to infer RV shifts}\label{sec:RV_extraction}

\subsubsection{\rev{Maximum-likelihood approach}}
\rev{If a photon with wavelength $\lambda$ is emitted by a target star with velocity $v$ along the line of sight from the Solar System barycentre, and negligible velocity perpendicular to the line of sight, relativistic Doppler stretching means it will be received with wavelength $\tilde{\lambda}$ by an observer in the Solar System
\begin{equation}
\tilde{\lambda} = \tilde{\lambda}(\lambda,v)  = \lambda \times \sqrt {\frac{{1 + v/c}}{{1 - v/c}}},
\end{equation}
where $c$ is the speed of light \citep{sher68}. Using this prescription, we can investigate which velocity $v$ will maximise the alignment between pairs of observed spectra. To formulate the joint GP likelihood $\mathcal{L}_{ij}$ for spectra $i$ and $j$, given a set of parameters including their relative velocity $v$, we make a simple modification to equation~\ref{eq:NLL_GP}:
\begin{align}
\log {{\cal L}_{ij}} &= p({{\cal S}_i},{\mathcal{S}_j}|v,{\boldsymbol{\theta} _{ij}},{\boldsymbol{\phi} _{ij}}) \nonumber \\ 
&=  - {\textstyle{1 \over 2}}\log \det {{\bf{K}}_{ij}} - {\textstyle{1 \over 2}}{{\bf{r}}_{ij}}^{\rm{T}}{{\bf{K}}_{ij}}^{ - 1}{{\bf{r}}_{ij}} - {\textstyle{M}}\log 2\pi ,
\label{eq:GP_NLL2}
\end{align}
where $\mathbf{r}_{ij}$ is a vector of residuals formed by subtracting the mean function evaluated at $\boldsymbol{\lambda}_{ij}(v)=[\boldsymbol{\lambda}_{i}^{\rm T}, \boldsymbol{\tilde{\lambda}}_{j}^{\rm T}]^{\rm T}$ from the observed fluxes $\mathbf{f}_{ij}=[\mathbf{f}_{i}^{\rm T}, \mathbf{f}_{j}^{\rm T}]^{\rm T}$, and $\mathbf{K}_{ij}\in\mathbb{R}^{2M\times 2M}$ is the covariance matrix whose elements are computed by evaluating equation~\ref{eq:goldilocks-kernel} for all pairs of wavelengths in $\boldsymbol{\lambda}_{ij}$. The value of $v$ that maximises the likelihood then corresponds to the best estimate of the RV shift between the two spectra:
\begin{equation}
\label{eq:RV-from-ML}
{\rm{R}}{{\rm{V}}_{ij}} = \mathop {\arg \max }\limits_{|v| < c} {{\cal L}_{ij}},
\end{equation}
assuming the hyper-parameters ${\boldsymbol{\theta} _{ij}}$ and ${\boldsymbol{\phi} _{ij}}$ are also at their ML values.}

\rev{The above approach to inferring pairwise RVs is simple enough conceptually, though potentially quite expensive computationally, since learning ${\rm{R}}{{\rm{V}}_{ij}}$ requires computation of $\mathcal{L}_{ij}$ for different values of $v$, ${\boldsymbol{\theta} _{ij}}$ and ${\boldsymbol{\phi} _{ij}}$ to find the global optimum. Fortunately, simple approximations can speed up computation significantly, usually with little accuracy penalty.}

\rev{For example, covariance and continuum (hyper-)parameters for different spectra may be learned prior to pairwise RV extraction. In practice, local values of $\boldsymbol{\theta}_i$ and $\boldsymbol{\theta}_j$ (e.g.\ within a single \'echelle order -- see Section~\ref{sec:sub-order}) will be consistent across almost all pairs of spectra, and indeed it makes sense to require pairs of spectra to share a common set of covariance hyper-parameters $\boldsymbol{\theta}_{ij}$, since we assume pairs of stellar spectra are locally equivalent up to a simple Doppler shift. The background polynomial parameters need not be consistent for different spectra, given e.g.\ different blaze corrections, though these parameters might still be learned prior to RV extraction: after all, the continuum parameters peculiar to spectrum $i$, $\boldsymbol{\phi}_i$ should not change each time we compare spectrum $i$ to various other spectra $j,k,\ldots\neq i$. Alternatively, covariance and continuum (hyper-)parameters may be learned separately for each spectrum, and the GP posterior predictive distributions for each spectrum evaluated on an extremely dense grid, uniformly spaced in log wavelength; inferring $\text{RV}_{ij}$ may then proceed by aligning $\mu_i^*$ and $\mu_j^*$ in the usual maximum likelihood sense, albeit ignoring covariance between fluxes in different spectra. An advantage of the latter approach is that covariance matrix inversion need not be repeated for each pairwise RV computation.}

\rev{With the high-SNR and high-resolution HARPS-N spectra we consider in Section~\ref{sec:real-data}, we found that making either of these approximations had negligible effect on the extracted RVs: the difference between RVs extracted using equation~\ref{eq:GP_NLL2} and either of the faster approximate approaches was always $<2$~\cmps, and usually $<1$~\cmps. This accuracy penalty might be more significant in cases of lower SNR or resolution, though we defer such considerations to future work.}

\rev{Approximations aside, we also note that the above approach to RV extraction is directly amenable to various improvements. Priors could be imposed on the parameters (to favour small velocities, for example), in which case RV estimates could be derived from posterior probability distributions. If maximising accuracy of RV extraction is more important than minimising computational overheads, the assumption that stellar spectra are identical up to RV shifts could be relaxed, and temporal covariance between spectra explicitly modelled, to account e.g.\ for stellar activity variability (see Section~\ref{sec:future}).}

\subsubsection{\rev{Cross-correlation approach}}

\rev{The `classical' approach to RV extraction has been to cross-correlate observed spectra with a template spectrum; as already noted in Section~\ref{sec:intro}, this was the approach used to discover the first exoplanets via Doppler spectroscopy, and even today it is used in the data reduction pipelines of various important spectrographs.}

\rev{When certain conditions hold, cross-correlation and ML-based approaches to aligning functions can be expected to yield equivalent results \citep[e.g.][]{zucker03,aigrain04}. However, in practice there are various reasons to favour the latter approach over the former. For example, it has been shown that cross-correlation approaches are more sensitive than ML approaches to chromatic atmospheric effects on barycentric corrections, which can induce RV errors \citep{blackman19}. It may also be shown that, under certain circumstances, the mathematical properties of cross-correlation can lead to small though systematic errors in RV estimates; ML RV estimates do not suffer from the same issues (Rajpaul~et al., \emph{in prep.}). In general, then, we would disavow a cross-correlation based approach to RV extraction.}

\rev{However, given the persisting prevalence of cross-correlation Doppler spectroscopy, and to enable comparison with industry-standard pipelines, we show in this paper how our approach to RV estimation fares using both pairwise ML and cross-correlation approaches.}

\rev{Given two spectra with GP posterior predictive mean functions $\mu_i^*$ and $\mu_j^*$, the CCF may be computed as follows:
\begin{equation}
{\phi _{ij}}(v) = \int {\mu _i^*\left( \lambda  \right)\mu _j^*\left( {\tilde \lambda (\lambda ,v)} \right)d\lambda },
\end{equation}
and the RV shift between the two spectra computed from the maximum of the CCF:
\begin{equation}
\label{eq:RV-from-CCF}
{\rm{R}}{{\rm{V}}_{ij}} = \mathop {\arg \max }\limits_{|v| < c} {\phi _{ij}}.
\end{equation}
In practice, the GP mean functions $\mu_i^*$ and $\mu_j^*$ may be evaluated on a very dense grid, say with $<0.1$~\mps\ resolution, and then an efficient FFT algorithm used to compute the CCF via numerical convolution.}

\subsubsection{Error estimation}

\rev{Regarding error estimation, we note that each GP posterior distribution is defined by a mean function and a covariance matrix which may be used to draw different, random realisations of the model spectrum in question. Thus, estimation of the variance in ${\text{RV}}_{ij}$, i.e.\ $\sigma \left( {{\text{RV}}_{ij}} \right)^2$, may be carried out via a straightforward Monte-Carlo repetitions of the calculations outlined in Section~\ref{sec:RV_extraction}, regardless of whether RVs are inferred via an ML or CCF-based approach. There is no need to devise any scheme for weighting the different parts of the observed or model spectra; the Monte-Carlo error propagation approach ensures that everything from photon noise and GP fitting or interpolation errors to the fact that absorption lines will contribute much more strongly to a CCF than quasi-continuum will be built into the $\sigma \left( {{\text{RV}}_{ij}} \right)$ terms, and exactly in proportion with the extent to which they increase the scatter in repeated RV estimates.

This uncertainty could be estimated approximately using e.g.\ a second-order Taylor expansion of the likelihood function or CCF around the $\arg\max$, though the Monte-Carlo approach doesn't rely on any such approximations, and thus yields reliable results even in cases of asymmetric or multi-modal likelihood functions/CCFs.  The Monte-Carlo approach may also be used to estimate the \emph{covariance} between different RV estimators.

A more Bayesian (though computationally-expensive) approach to quantifying RV uncertainty would be to sample from the posterior distributions obtainable by combining the likelihood in equation~\ref{eq:GP_NLL2} with suitable priors. As this approach has no direct equivalent for CCFs, we defer its consideration to future work.}


 \subsection{RV extraction on an order-by-order (or finer) basis}\label{sec:sub-order}

In practice, it turns out to be important to model spectra in chunks, rather than all at once, for a few reasons. 

First, using a GP to model $M>200,000$~fluxes per spectrum is neither straightforward, given the $\sim\mathcal{O}(M^3)$~scaling of ordinary GP regression, nor necessary, given that we expect no covariance between widely-spaced spectral lines. Actually, for typical values of $\rho\sim0.25$~\AA\ on HARPS FGK dwarf spectra, covariances will decay to zero for separations spanning more than several per cent of a single order (typical wavelength span $\sim55$~\AA). This means that sparse matrix inversion algorithms may be leveraged to speed up the GP regression by many orders of magnitude -- with computational requirements approaching $\mathcal{O}(M)$ rather than $\mathcal{O}(M^3)$ scaling  -- while not sacrificing any accuracy \citep{reece2010e,sarkka2013,reece2014,dfm17}.

Second, given that typical spectral features and continuum behaviour may be expected to vary with wavelength, it will also be preferable to allow for different sets of hyperparameters for different parts of a spectrum, rather than forcing a single set of hyperparameters to model a single spectrum globally.

Third, and most important, not all spectral regions will encode the same RV information. \rev{As noted in Section~\ref{sec:pairwise-xcorr1}, while all stellar lines should undergo the same Doppler shifting as a result of reflex motion due e.g.\ to orbiting planets, stellar lines strongly distorted by an activity process will be measured to have a different RV than lines not affected by activity. Similar considerations would apply in the case of telluric contamination. Therefore it would be desirable to obtain a large ensemble of RV estimates for every spectrum, and to use the ensemble to try to figure out which may be unreliable indicators of the overall stellar Doppler shift}.

Therefore, let us assume that we will model any given spectrum in $L$ separate `chunks'; we shall denote by ${\rm{RV}}_{ij}^{(l)}$ the RV inferred by comparing the $l$\th\ sub-divisions of spectra $i$ and $j$. The calculations in Sections~\ref{sec:RV_extraction} proceed as before, except over restricted wavelength ranges indexed by $l$.

There are two salient choices we shall explore for the value of $L$. The first would be to set $L$ equal to the number of \'echelle orders in a given spectrograph. For HARPS-N spectra, this would mean $L=69$. This choice would enable a direct comparison with standard approaches to RV extraction with \'echelle spectra, wherein all orders are essentially treated as independent spectra, with the RVs for each order being combined in some manner to produce a final RV. But this value of $L$ arises from properties of the \'echelle grating and not the spectrum itself, and we can expect to do better with a larger value of $L$. If every order contains at least some degree of activity or telluric contamination, then the RV estimates for all order will be biased by this contamination to a lesser or greater extent; the larger variance associated with more strongly contaminated orders \emph{should} dilute the effect in the final variance-weighted averaging, but their effect will remain non-zero. 

\rev{If instead} we make $L$ large enough so that each chunk or sub-order (the term we shall henceforth use for chunks smaller than an order) only contains anything between a fraction of a line and a few spectral lines\footnote{To keep our technique as general as possible, we assume that we have no no prior knowledge of line locations or morphologies.} we should be able to extract differential RVs on a quasi line-by-line basis, and resolve signatures of localized activity or telluric contamination, if we find that RVs extracted from a particular sub-order often diverge from the global set of RVs. Because such contamination should have nothing to do with the true dynamical stellar RV, we could exclude these regions of spectra  from the final RV calculation, and check whether the scatter in the extracted RVs increases or decreases.\footnote{This approach would not be able to identify hypothetical non-localized contamination that affected \emph{all} spectral lines.}  The Calcium {\sc ii} H \& K lines would be an extreme example of where we would find such contamination, though in principle we could identify other, far weaker contamination that was not known \emph{a priori}. Ideally $L$ would be as large as possible while still ensuring that sub-orders were wider than several covariance length scales, and also wide enough so that even the largest expected RV shifts (e.g.\ $\pm30$~\kmps\ for telluric features after barycentric velocity correction) could not result in the same two sub-orders from different spectra having no spectral features in common.

\subsection{Aggregating the pairwise RV shifts}
With the ${\text{RV}}_{ij}^{(l)}$ \rev{arrays} computed -- \rev{whether via cross-correlation or, preferably, an ML approach} -- our final task is to aggregate the pairwise RV shifts to estimate a single RV shift for each spectrum.

Following the idea sketched in Section~\ref{sec:pairwise-xcorr2}, we can average over the pairwise RV shifts in each column (or row) to estimate an RV shift for the $l$\th order in the $i$\th\ spectrum, relative to some arbitrary offset. We compute a variance-weighted mean RV as follows:

\begin{equation}\label{eq:avg1}
{\rm{RV}}_i^{(l)} = \sum\nolimits_{j = 1}^N {w_{ij}^{(l)}{\rm{RV}}_{ij}^{(l)}}  + \gamma,
\end{equation}
with the variance weights \citep[cf.][]{schmelling95} given by
\begin{equation}\label{eq:avg2}
{{w_{ij}^{(l)} = \sigma {{\left( {{\rm{RV}}_{ij}^{(l)}} \right)}^{ - 2}}} \mathord{\left/
 {\vphantom {{w_{ij}^{(l)} = \sigma {{\left( {{\rm{RV}}_{ij}^{(l)}} \right)}^{ - 2}}} {\sum\nolimits_{k = 1}^N {\sigma {{\left( {{\rm{RV}}_{ik}^{(l)}} \right)}^{ - 2}}.} }}} \right.
 \kern-\nulldelimiterspace} {\sum\nolimits_{k = 1}^N {\sigma {{\left( {{\rm{RV}}_{ik}^{(l)}} \right)}^{ - 2}}.} }}
\end{equation}
As $\gamma$ is a constant for all $i$, we can set $\gamma=0$.  The variance in this weighted mean may be computed as
\begin{eqnarray}\label{eq:avg3}
\sigma {\left( {{\rm{RV}}_i^{(l)}} \right)^2} & = & {\mathop{\rm cov}} \left( {{\rm{RV}}_i^{(l)},{\rm{RV}}_i^{(l)}} \right) \nonumber \\
& = & \sum\nolimits_{j,k = 1}^N {w_{ij}^{(l)}w_{ik}^{(l)} \cdot {\mathop{\rm cov}} \left( {{\rm{RV}}_{ij}^{(l)},{\rm{RV}}_{ik}^{(l)}} \right)}.
\end{eqnarray}
The limiting $1/\sqrt{N}$ behaviour of the variance in the weighted mean is easily verified for the simple case where the RV estimators are uncorrelated, so that ${{\mathop{\rm cov}} \left( {{\rm{RV}}_{ij}^{(l)},{\rm{RV}}_{ik}^{(l)}} \right)}=0\;\forall\;i\neq j$, and all the $\sigma {\left( {{\rm{RV}}_{ij}^{(l)}} \right)^2}$ are equal. The presence of non-zero covariances between different RV estimators will tend to increase the variance in the weighted mean, since the different estimators no longer provide completely independent information on the overall value. Note that the value assigned to $\gamma$ will not affect the variance, which in general is invariant with respect to a location parameter \citep{wasserman04}.  

We can now also compute a variance-weighted mean across all $L$ \'echelle orders or sub-orders to obtain a final estimate of the stellar RV of the $i$\th\ spectrum:

\begin{equation}\label{eq:avg4}
{\rm{R}}{{\rm{V}}_i} = \sum\nolimits_{l = 1}^L {w_i^{(l)}{\rm{RV}}_i^{(l)}},
\end{equation}
with the variance weights given by
\begin{equation}\label{eq:avg5}
w_i^{(l)} = {{\sigma {{\left( {{\rm{RV}}_i^{(l)}} \right)}^{ - 2}}} \mathord{\left/
 {\vphantom {{\sigma {{\left( {{\rm{RV}}_i^{(l)}} \right)}^{ - 2}}} {\sum\nolimits_{m = 1}^L {\sigma {{\left( {{\rm{RV}}_i^{(m)}} \right)}^{ - 2}}} }}} \right.
 \kern-\nulldelimiterspace} {\sum\nolimits_{m = 1}^L {\sigma {{\left( {{\rm{RV}}_i^{(m)}} \right)}^{ - 2}}} }},
\end{equation}
and the variance in the final RV estimate given by
\begin{eqnarray}\label{eq:avg6}
\sigma {\left( {{\rm{R}}{{\rm{V}}_i}} \right)^2} &=& \sum\nolimits_{l,m = 1}^L {w_i^{(l)}w_i^{(m)} \cdot {\mathop{\rm cov}} \left( {{\rm{RV}}_i^{(l)},{\rm{RV}}_i^{(m)}} \right)}  \nonumber \\
&=& {1 \mathord{\left/
		{\vphantom {1 {\sum\nolimits_{m = 1}^L {\sigma {{\left( {{\rm{RV}}_i^{(m)}} \right)}^{ - 2}}} }}} \right.
		\kern-\nulldelimiterspace} {\sum\nolimits_{m = 1}^L {\sigma {{\left( {{\rm{RV}}_i^{(m)}} \right)}^{ - 2}}} }};
\end{eqnarray}
the final expression in equation~\ref{eq:avg6} is simpler than that in equation~\ref{eq:avg3} because we know \emph{a priori} that within a given observation, a given (sub-)order should provide an RV constraint independent of those from other orders containing different spectral features. If we identify certain sub-orders as being contaminated by activity or the like -- and we shall soon consider simple strategies for doing this -- the summations above would be adjusted to exclude the sub-orders in question.


To recap, we started with a three-dimensional, $N\times N \times L$ array of RV estimates, then reduced it to a two-dimensional $N \times L$ matrix, which may be thought of as a set of $N$ differential RVs for each $l=1,2,\ldots,L$ (sub-)order. Finally, we reduced this to a set of $N$ differential RVs. 


\section{Tests with synthetic data}\label{sec:synth-data}
\subsection{Overview of tests and synthetic spectra}

To ascertain that our method worked as intended, we ran a number of tests using simple, synthetic spectra. We present here a representative subset of these tests.

\begin{figure}
	\begin{center}
		\includegraphics[width=\columnwidth]{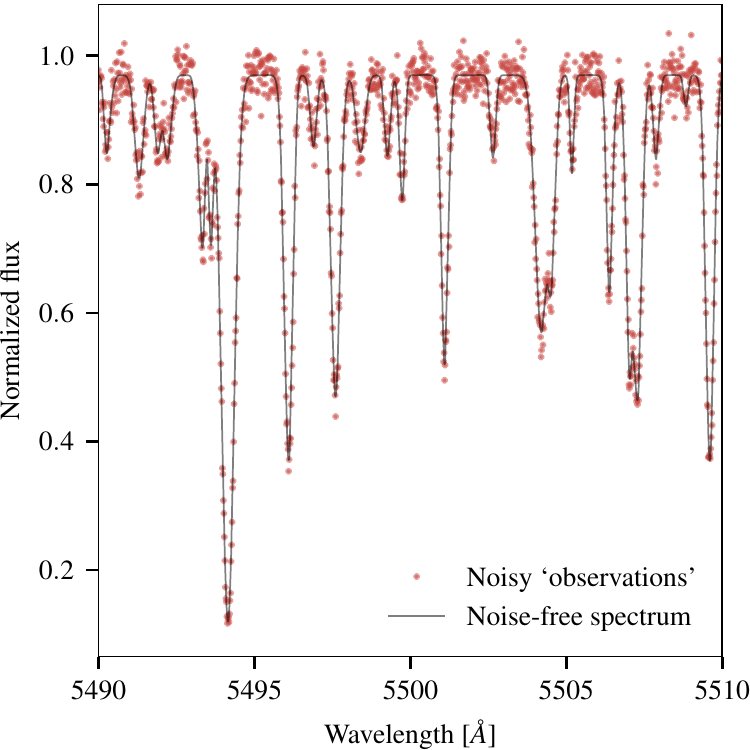}
		\caption[]{Representative example of a portion of one of our synthetic spectra, covering about a third of an \'echelle order, with $R=100\;000$ and $\text{SNR}=50$.}
		\label{fig:synth_spec_example}
	\end{center}
\end{figure}

The synthetic spectra in the tests described below comprised $69$~\'echelle~orders with the same wavelength coverage as HARPS-N spectra, where the spectrum in each order was synthesized as a forest of Gaussian absorption lines on a flat continuum, convolved with a Gaussian profile with FWHM$=\lambda/R$ , and with Poisson noise of varying levels added to each synthetic observation. The individual absorption lines had randomly-chosen parameters, but with the overall number and shapes of lines chosen so that in the $R\sim100\;000$ case, the Mat\'ern covariance hyper-parameters for a GP model fitted to that synthetic \'echelle order would match the hyper-parameters derived from the corresponding \'echelle order from a real HARPS-N spectrum, specifically from a G5\;V dwarf (cf.\ Section~\ref{sec:real-data}). A portion of one such synthetic spectrum is shown in Fig.~\ref{fig:synth_spec_example}.

While our synthetic spectra do bear superficial resemblance to real stellar spectra, they are of course far too simplistic to be considered realistic. However, this was by design. We wanted to be able to control and vary all inputs to our method, and to be able to study cases both with and without any `confounding' factors present, i.e. anything more complicated than genuine stellar RV shifts, e.g.\ activity-like contamination. Apart from this, as our method is completely agnostic to the actual shapes of the spectra being fitted, it should work just as well if including (say) more realistic Voigt or Lorentzian line profiles. Therefore we favoured simplicity in these proof-of-concept tests.

\subsection{Basic tests}


In our first set of tests, we injected into $4096$ synthetic spectra with epochs spanning one year a year a zero-eccentricity Keplerian signal with period $35$~d and semi-amplitude $10$~\mps, which was chosen to be large enough so that it could be recovered even in cases of relatively low SNR and spectrograph resolution. Apart from the time-dependent RV shifts and barycentric velocity corrections, and different noise realisations, the spectra were identical. We then studied the accuracy and precision with which we could use our method to recover the injected signal, as a function of varying spectral SNR, resolution, and number of spectra modelled ($N=2,3,\ldots,4096$). To compare our injected and extracted differential RVs, we constrained both sets to have the same variance-weighted mean $\gamma$, which without loss of generality we set to zero (see equation \ref{eq:gamma-origin}). For comparison, we also extracted RVs by cross-correlating each synthetic spectrum with the noise-free, infinite resolution template from which the synthetic observations were generated.

Our finding was that regardless of SNR or spectrograph resolution, errors in extracted RVs did indeed exhibit a $\sim1/\sqrt{N}$ behaviour (allowing for random-walk like variations), up to a certain point, after which the relation flattened out. In the limit of large $N$, we found that the errors in the RVs extracted using our technique became equivalent (within insignificant numerical fluctuations) to the errors associated with the idealised case where the perfect template was available. 

Both the initial errors and the asymptotic error behaviour for large $N$ depended on SNR and resolution of the synthetic spectra, with the dependence being quite straightforward in the case of SNR: doubling SNR, as may be achieved in practice by increasing integration times by a factor of four, led to a halving of errors. The dependence was somewhat less straightforward in the case of resolution, as different lines became (un)resolved at different resolutions \citep{bouchy01}. 

\begin{figure*}
	\begin{center}
		\includegraphics[width=\textwidth]{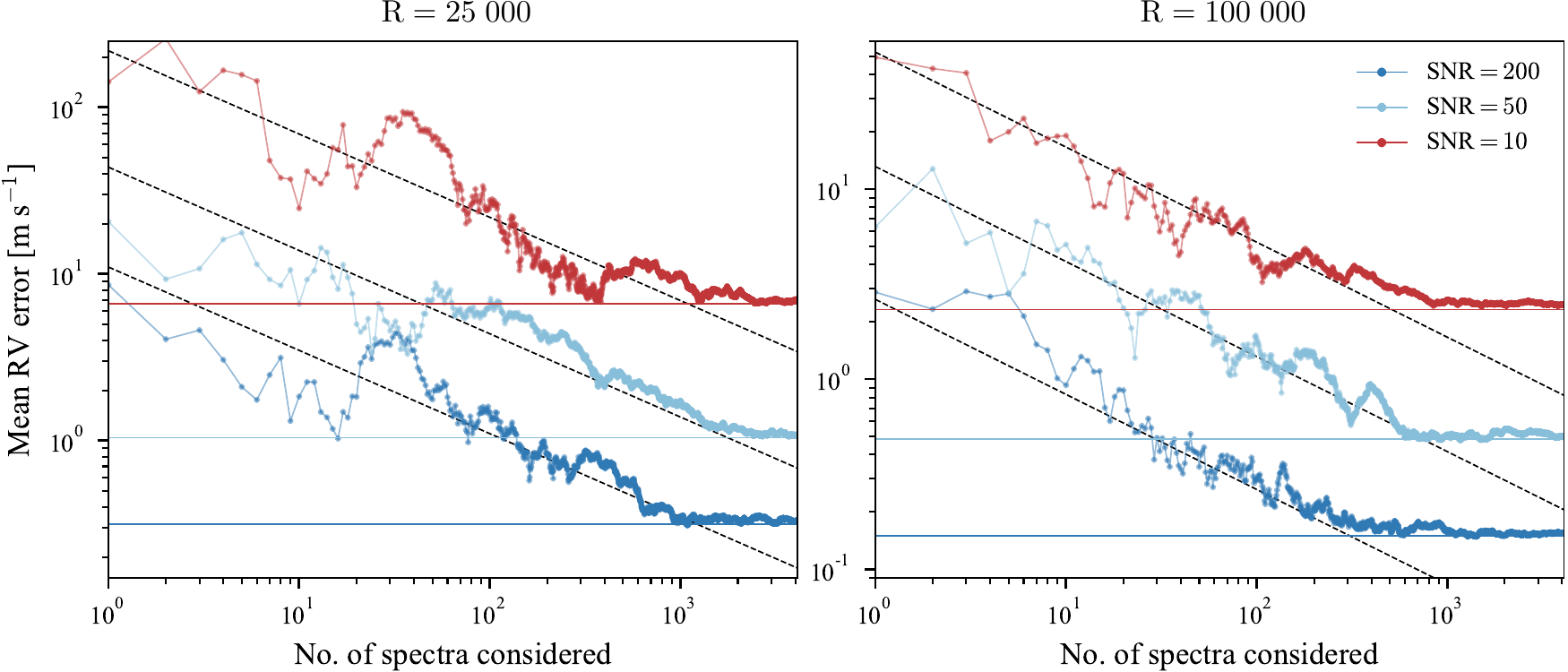}
		\caption[]{Errors in RVs extracted from synthetic spectra using our new technique as a function of the number of spectra modelled ($N$), for different resolutions and SNRs. The errors were calculated by taking the mean absolute difference between extracted and injected RVs. The dotted black lines have slope $-\tfrac{1}{2}$, and are included as a visual aid to highlight the $\sqrt{1/N}$ scaling of errors. The \emph{estimated} errors (not plotted here) agreed very closely with the actual errors in the sense that the differences between estimated and true RVs were smaller than the formal $1\sigma$ error estimates $\sim70\%$ of the time. The horizontal lines indicate the mean errors obtained when a perfect template was available. In these tests, then, our technique performed as expected both in terms of accuracy and precision.}
		\label{fig:synth_01}
	\end{center}
\end{figure*}

A few representative results are presented in Fig.~\ref{fig:synth_01}. In the cases shown there, the GP errors are indistinguishable from the perfect template case after about $1000$ spectra are available. Given how the observations were synthesised, we might expect roughly similar considerations to apply in the case of a real, slowly-rotating, Sun-like star observed with a moderate to high-resolution spectrograph. However, in general -- for different stellar types, rotation velocities, spectrograph resolutions, observing calendars, etc.\ -- simple numerical simulations involving at least a couple of real spectra would be useful to estimate the number of observations required for the Cram\'er-Rao lower bound on RV variance to be achieved, i.e.\  for the point to be reached at which no more improvements to RV precision are possible.

It is difficult to compare on an even footing the performance of our technique with alternative approaches to RV extraction, in large part because of the many free parameters and assumptions that would need to make when implementing any such alternative approach (see however Section~\ref{sec:future}). For instance, if we wanted to use a binary mask, RV extraction performance would be sensitive to assumptions about errors in the locations of the lines incorporated into the mask, the fraction of lines we neglected, the extent of template mismatch between the mask and the observations, etc. If we were to use some sort of scheme for direct construction of a template from the observations, performance would be sensitive to the iterative scheme we used for un-shifting and stacking observations, how we interpolated observations to common wavelength scales (e.g.\ linear vs.\ polynomial vs.\ spline interpolation), how we estimated errors, etc. 

A simple way to get a handle on the possible advantages specifically of the GP \rev{(though not the pairwise RV extraction)} component of our model was to replace our Mat\'ern covariance kernel with a white noise kernel, such that all fluxes in a spectrum were treated as independent, but to leave all other parts of our technique unchanged. We found that when doing so, the $\sqrt{1/N}$ error behaviour was still observed, but the errors for small $N$ were larger than when not neglecting covariance over wavelengths, such that it took longer for errors to decrease below a given threshold. The advantages of modelling covariances were greatest in cases of low resolution and SNR, with errors being up to a few $\mps$ smaller for small to moderate $N$. 
In cases of very high resolution and SNR (e.g.\ $R=100\;000$, $\text{SNR}\sim200$) the advantages of the Mat\'ern kernel became negligible, since the observations themselves were already close to being noise-free realisations of the underlying spectrum. 

\subsection{Adding activity- and telluric-like contamination}

\begin{figure}
	\begin{center}
		\includegraphics[width=\columnwidth]{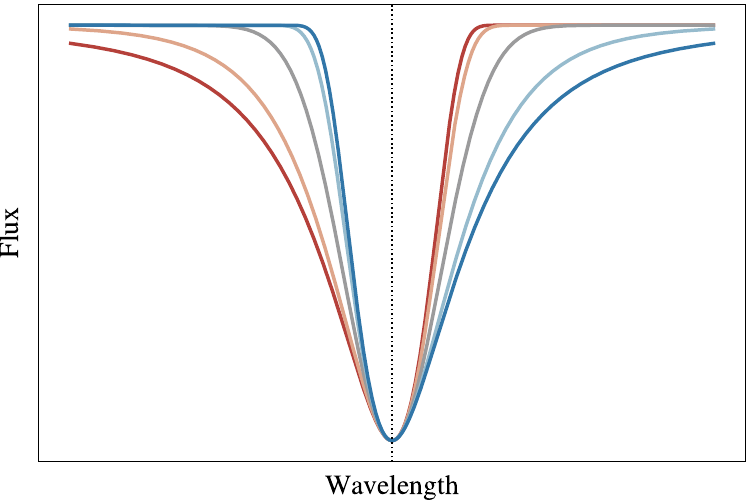}
		\caption[]{Illustration of the type of variability we introduced into a subset of Gaussian spectral lines, for five different phases of the time-varying asymmetry, in all cases with $\sigma_1/\sigma_0=5$ (chosen so that the asymmetry may be easily discerned visually) and zero relative Doppler shift.}
		\label{fig:skew_gauss}
	\end{center}
\end{figure}

To simulate contamination due e.g.\ due rotating active regions, we introduced a time-varying asymmetry into a randomly-chosen subset of lines in our synthetic spectra. For lines that originally had the form $\ln f = \frac{{ - {{(x - \mu )}^2}}}{{2\sigma _o^2}} + f_0$, we introduced an asymmetry by letting the line width depend both on time and on the position within the line: ${\sigma _0} \to \sigma (x,t) = {\sigma _0} - {\sigma _1}(x - \mu )\cos \tfrac{{2\pi t}}{T}$, where $\sigma_1 < \sigma_0$ is a skewness parameter, and $T$ defines the line asymmetry oscillation period. The net effect of this parametrization was to introduce into spectra an apparent though spurious quasi-sinusoidal RV shift with period $T$, as might be expected from rotating active regions; see Fig.~\ref{fig:skew_gauss}. The skewness parameters and number of stellar lines affected were varied in order to achieve a desired level of activity-like contamination (e.g.\ a $10$~\mps\ signal with $25$~d period).

Additionally, to simulate the effect of telluric-like contamination, we imprinted on each synthetic spectrum a second spectrum with zero overall RV relative to our imagined observatory, such that after barycentric correction this `telluric' spectrum would shift by $\pm30~\kmps$ over the course of a year. To simulate some orders being much more strongly contaminated by telluric features than others, we enforced the arbitrary constraint that only a quarter of the \'echelle orders would contain any telluric features. The synthetic telluric spectrum was synthesised using only Gaussian lines, whose parameters were randomly varied to achieve a desired level of contamination (e.g.\ $1$~\mps) when incorrectly treating the telluric spectrum as part of the stellar spectrum.

To make these observations more realistic than in the first set of tests, we now considered only $500$ observations taken over the course of three and a half years, with gaps in the observations based on a real HARPS-N observing schedule, and we reduced the SNR of the stellar spectrum to $50$. We left the properties of the injected Keplerian signal unchanged, and fixed $R=100\;000$. We experimented with a range of activity and telluric contamination levels, though for the sake of brevity present in the next section the results from just one representative test.

\subsection{Simple approaches to mitigating the contamination}\label{sec:reduce_contamination}

Using synthetic data containing only activity-like contamination, only telluric-like contamination, and both, we found that it was useful to extract a separate RV signal from each sub-order, and then to compute the following contamination-proxy statistics:
\begin{enumerate}
	\item the mean uncertainty in the RVs extracted for that sub-order;
	\item the rms scatter in the RVs extracted for that sub-order; and
	\item the mean linear correlation between the RVs extracted for that sub-order, and the RVs extracted using all sub-orders.
\end{enumerate}
By masking off orders with extreme (outlying) values of these statistics,\footnote{\rev{In Section~\ref{sec:future} we suggest more sophisticated possible approaches to identifying spectral contamination.}} we were usually able to diminish very significantly the telluric- and/or activity-like contamination in the final extracted RVs. 

The first of the above three statistics, when unusually large, identifies regions of spectra where RVs could not be inferred as precisely as elsewhere, e.g.\ because of genuine Doppler signals being diluted by activity and/or telluric variability, \rev{or simply because the regions are dominated by continuum emission rather than stellar absorption lines}. The second statistic, when unusually large, identifies regions where the RVs themselves vary more over time, regardless of how precisely they could be constrained, as might be expected e.g.\ from telluric lines moving with a high velocity due to Earth's barycentric motion, or from a strong activity signal. The third statistic, when unusually small or even negative, identifies regions of spectra which do not, on average, shift in the same way as most of the rest of the spectrum -- as should be the case for both mild to moderate activity and telluric contamination (barring exceptional and hypothetical cases where the phase and the period of contaminant matched that of the Doppler signal). While we may be examining very small segments of spectra, and will thus be dealing with relatively weak signals, we make up for it to some extent by aggregating information across the time dimension; the more spectra we have available, the more constraining these statistics should be.

Once computed, we can then systematically check whether excluding sub-orders with the most unfavourable values for these statistics decreases the rms scatter in the extracted RVs; in practice the three statistics will usually be significantly correlated but not degenerate. We expect the rms to decrease whenever we exclude genuine contamination, as the scatter of a Doppler signal plus either activity or telluric variability should be greater than the scatter of the Doppler signal alone, except in the pathological and unlikely case where the contaminating red noise happens to have a very similar period to and to be in anti-phase with the Doppler signal. On the other hand, if we throw away useful regions of spectra that would otherwise have helped constrain the Doppler signal, we'd expect the rms scatter to \emph{increase} due to greater white noise in the extracted signal.  
To minimise assumptions made about the contamination, sub-orders might be excluded non-parametrically (e.g.\ exclude the $1,2,3,\ldots$ sub-orders with the lowest ranked mean correlation with other spectra), rather than via a scheme such as $\sigma$-clipping, which would necessitate an assumption of a symmetric and short-tailed distribution for the statistic in question. 
Finally, we should note that we are assuming that any contamination present is relatively localized: for hypothetical contamination that dominated the Doppler signal in more than half of the spectrum, our masking approach could lead to the Doppler signal being rejected in favour of the contaminant. Even in such a case, though, two separate Doppler signals could still be extracted; one would then need to figure out which to ascribe to stellar reflex motion and which to contamination.  


To illustrate all of this in practice, we show results here for a representative example where we set the `activity' period to be $T=25$~d, and randomly chose the numbers of lines affected and their skewness parameters in such away that the net effect of the contaminated lines' presence (when fitting all orders in all spectra without masking, though performing the usual variance-weighting etc.) would be an apparent $\sim10$~\mps\ signal in the extracted RVs, i.e.\ of same order as the Keplerian signal's semi-amplitude, while the impact of the `telluric spectrum' on the final RVs would by itself be of order $2$~\mps\ \citep[see][]{cunha14}.

We then investigated splitting each order into an integer number of sub-orders, $n_\text{split}$, and extracting an RV signal for each sub-order, instead of for each full order. By iteratively excluding certain sub-orders from the computations to produce the final set of RVs, particularly those with unfavourable values of the contamination `proxies' suggested above, we were able to reduce the rms scatter in the final RV signal from $14.6$~\mps\ to $10.6$~\mps, after which point the rms scatter started increasing as we masked off more sub-orders. We obtained best results with $n_\text{split}=16$ and masking off approximately two-fifths of all sub-orders in this particular test, although these values reflect various inputs to our synthetic spectra, and should be expected to be (possibly very) different for real spectra; we consider in some detail the particulars of applying this process to a real data set in Section~\ref{sec:reducing_rms}. In any case, the principle at play here is that all else being equal, throwing away photons by excluding sub-orders should ordinarily lead to an \emph{increase} in RV scatter due to increased white noise variability in the RVs. However, if the scatter actually \emph{decreases}, then we can infer that we have removed an even greater degree of red/correlated noise from the RVs, to compensate for the inevitable increase in white noise variability.

\begin{figure}
	\begin{center}
		\includegraphics[width=\columnwidth]{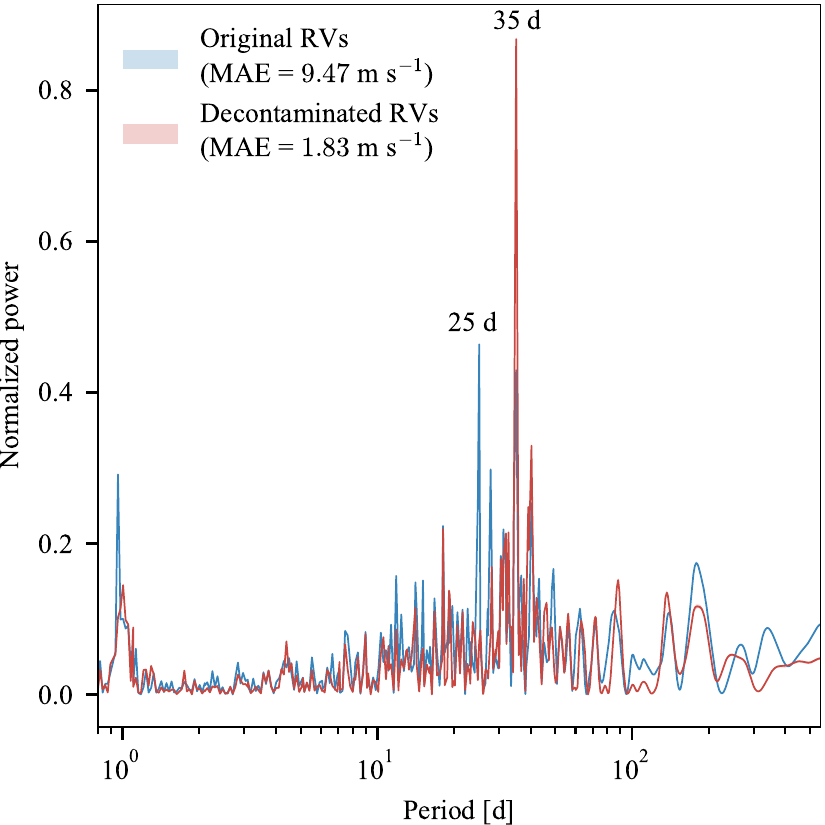}
		\caption[]{Lomb-Scargle power spectra of RVs extracted from a set of synthetic spectra, before and after excluding certain sub-orders to reduce activity- and telluric-like contamination, along with the mean absolute errors (MAE) for each set of RVs. The $25$~d periodicity due to activity-like contamination is strongly suppressed in the decontaminated RVs, while the power of the $35$~d Keplerian period is boosted.}
			\label{fig:lomb_scargle0}
	\end{center}
\end{figure}

In Fig.~\ref{fig:lomb_scargle0}, we plot the power spectrum of RVs extracted with and without this data-driven masking designed to mitigate contamination. We see that in the former case, the $25$~d activity period shows up very prominently, as does the $35$~d Keplerian period, and in fact the mean absolute RV error turns out to be $9.47$~\mps, due largely to the very strong activity contamination. With masking, the $25$~d period is greatly suppressed, as is some variability on much longer time scales, while the $35$~d signal is boosted; the mean absolute RV error is now $1.83$~\mps: a fivefold improvement, though still some way off the $0.70$~\mps\ that was obtainable in the contamination-free case, indicating that our simplistic method managed to localize and mitigate most though not all of the activity- and telluric-like contamination.

While such tests with synthetic data are far from realistic, they do at least serve to establish proof-of-concept: given spurious RV variability from line shape distortions, a superimposed spectrum that shifts with the Earth's barycentric velocity rather than the RV common to all stellar lines, etc., simple modifications to our basic technique can be used to identify and mitigate such contamination. Of course, the approach could be made much more sophisticated. For example, one might use a machine learning approach to choose some linear combination of RVs from different sub-orders that would minimize the scatter in the extracted RVs, rather than relying on a few crude proxies to choose sub-orders for exclusion. Better still, one could explicitly model time-varying changes in the observed spectra (see Section \ref{sec:future}), which could obviate the need for any masking at all. 

Whatever the shortcomings of our synthetic data, a far more valuable validation of our technique comes in the \rev{next two sections}, where we show that it can be applied to a real data set and yield better results than \rev{widely-used RV extraction pipelines}.

\section{Na\"ive application to a real data set}\label{sec:real-data}
\begin{figure*}
	\begin{center}
		\includegraphics[width=\textwidth]{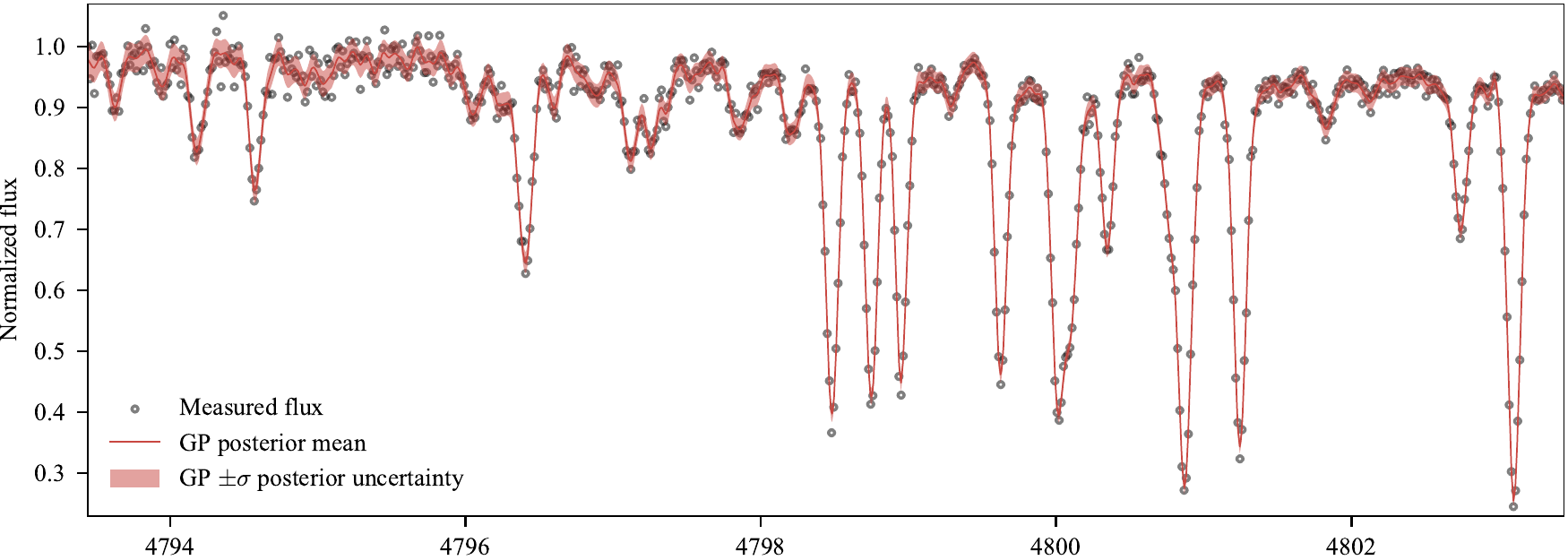}
		\includegraphics[width=\textwidth]{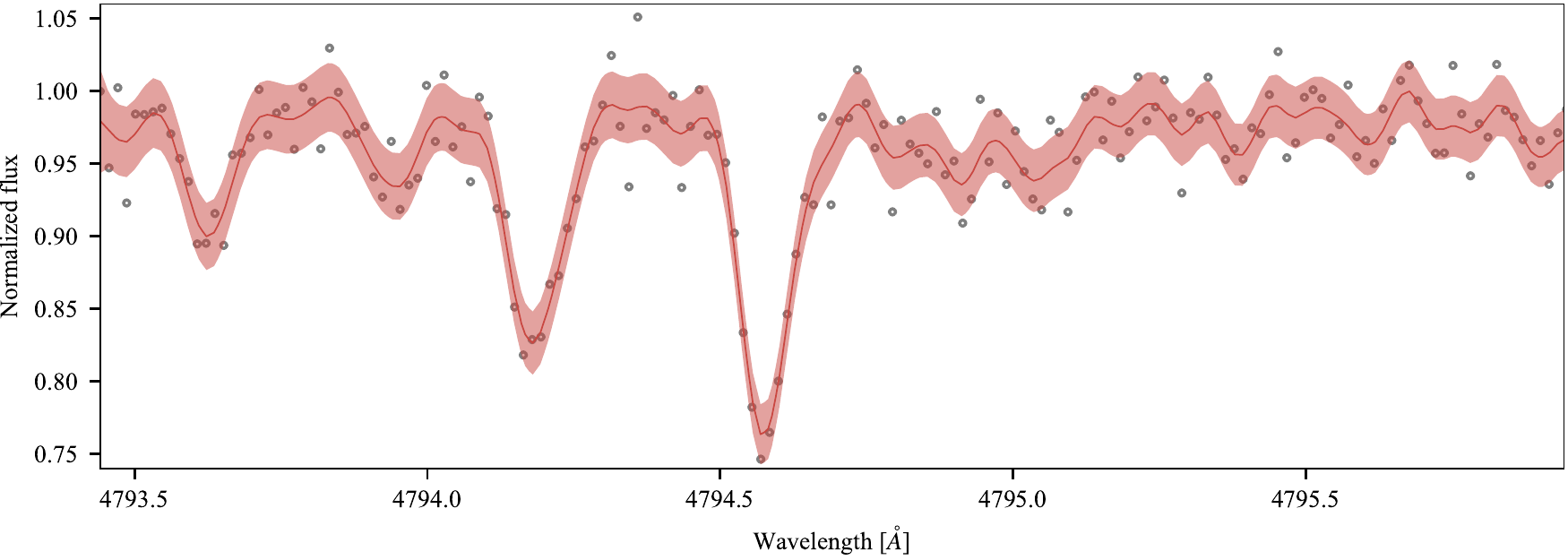}
		\caption[]{Part of a typical GP fit to a randomly-selected \mystar\ spectrum. The wavelength coverage in the top panel corresponds to about $20\%$ of one \'echelle order, while the lower panel zooms in to cover only $5\%$ of the same order. To minimize clutter and make the plots easier to inspect, the posterior mean has been re-normalized to the unit interval, and error bars on the measured fluxes (proportional to the square root of the fluxes) have been suppressed.} 
		\label{fig:GP_RV_fit}
	\end{center}
\end{figure*}

\begin{figure*}
	\begin{center}
		\includegraphics[width=\textwidth]{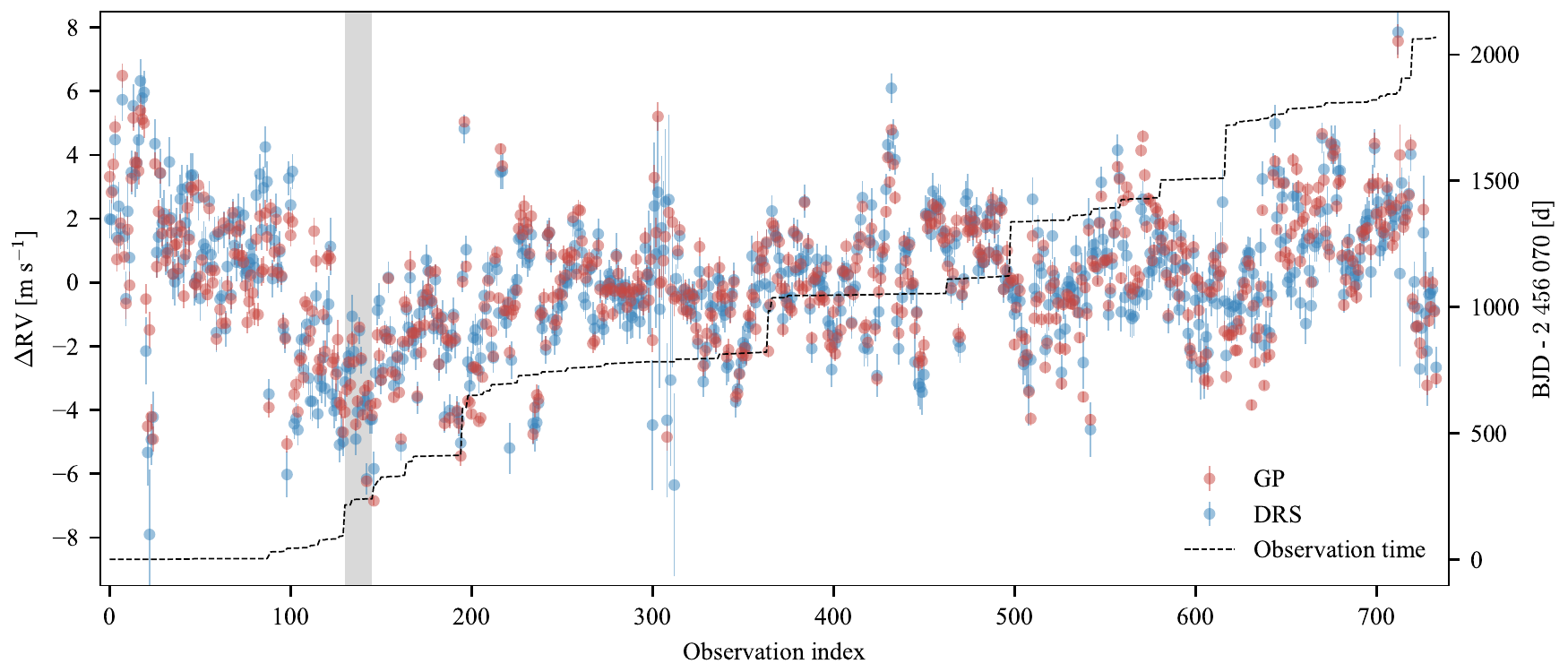}
		\caption[]{RVs extracted using our GP-based technique (red) and the DRS pipeline (blue), along with $1\sigma$ error bars. Because the time sampling alternates between very dense coverage (multiple observations per night), one or two observations per night, and very large gaps between observing seasons, the RVs are plotted as a function of serial observation \emph{index} ($0,1,2,\ldots,734$) rather than time; the time corresponding to a given observation may be read off from the dashed line's position on the right-hand vertical axis. This representation obviates the need for many different plots with different limits for the time axes. The greyed-out region denotes the period between the failure of HARPS-N's original CCD in September 2018, and the end of the commissioning of the replacement CCD in February 2013.}
		\label{fig:RV_serial}
	\end{center}
\end{figure*}
\begin{figure*}
	\begin{center}
		\includegraphics[width=\textwidth]{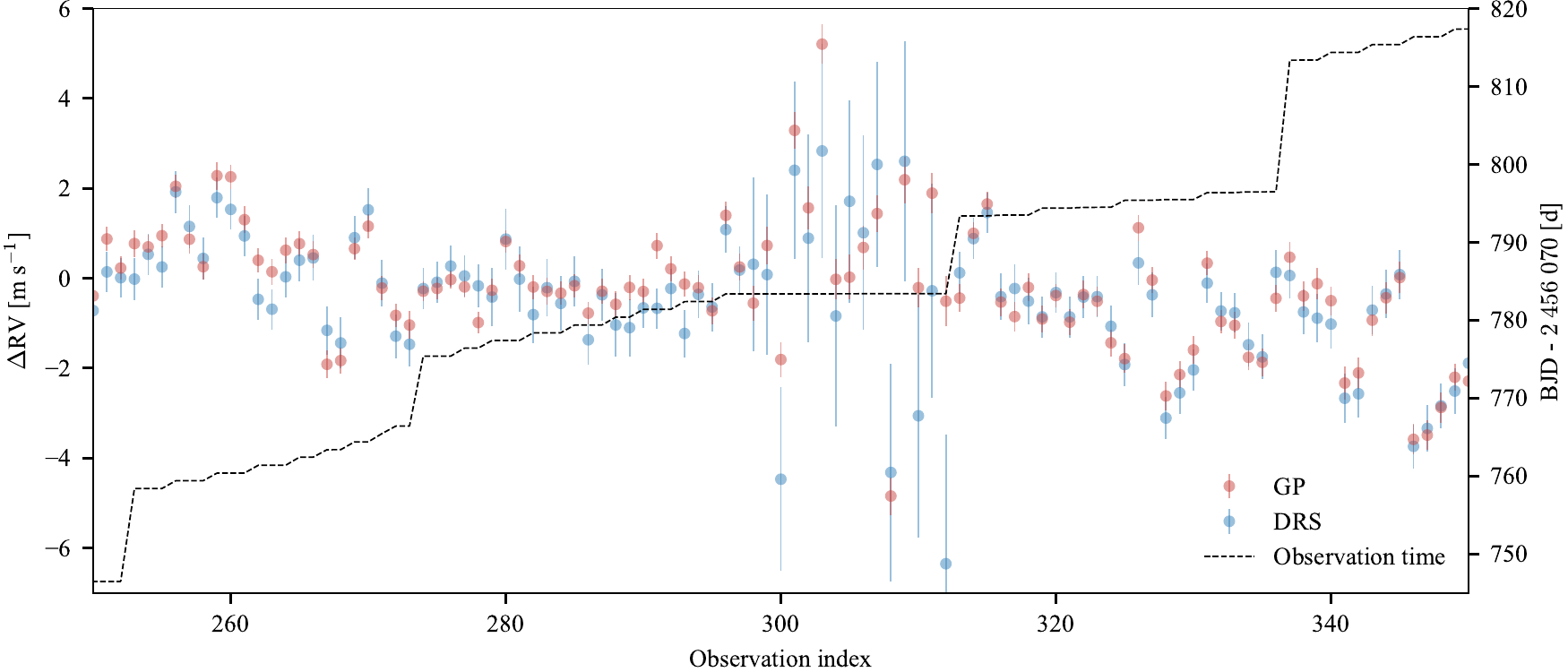}
		\caption[]{As for Fig.~\ref{fig:RV_serial}, but zoomed in to cover about $90$ observations made over a baseline of about $10$ weeks. The rms scatter of the GP RVs is very slightly though noticeably smaller for observations more closely spaced in time. The GP RV extraction technique, as implemented here, ignores the observation times, and so has no way of `knowing' whether observations are closely or widely spaced in time (see Section~\ref{sec:future} for a discussion on how this temporal information could be used to refine RV estimates). Observations $310$ to $324$ had integration times of just $15$~s, so their increased RV scatter is probably due to an increase in both white \emph{and} red noise, associated with a decrease in photons and in p-mode averaging (see Section~\ref{sec:GP_vs_N}), respectively.}
		\label{fig:RV_serial2}
	\end{center}
\end{figure*}
\rev{In Sections~\ref{sec:mystar} to \ref{sec:GP_vs_N} below, we use our GP technique to derive RVs from spectra of the standard star \mystar; we focus on demonstrating a high level of consistency with RVs derived from the `industry-standard' DRS (Data Reduction Software) pipeline used on both HARPS and HARPS-N. To this end, we align two aspects of our overall technique with the latter pipeline. First, we infer RVs via cross-correlation rather than ML estimation (i.e.\ we use equation~\ref{eq:RV-from-CCF} rather than equation~\ref{eq:RV-from-ML}); and second, we estimate the location of CCF peaks via Gaussian fits to the points bracketing the peak, which we are informed is the approach used by the DRS.}\footnote{\rev{The theoretical justification for the Gaussian approximation is unclear. A Taylor expansion around the global maximum of an arbitrary function is parabolic, to second order, with the linear term vanishing. Exponentiating a Taylor approximation of the \emph{logarithm} of function does lead to a Gaussian approximation to the function (as e.g.\ with log likelihoods), though this would not be true for an arbitrary, non-Gaussian CCF. Tests with synthetic spectra suggested parabolic fits lead to marginally more accurate RVs.}} 


Even though we apply our technique na\"ively by blindly fitting all spectra and all orders ($L=69$),\footnote{We do however exclude from the RV calculations, per equations \ref{eq:avg2} to \ref{eq:avg6}, a small fraction of pairwise RVs with uncertainties at least an order of magnitude larger than the median uncertainty, indicating e.g.\ imperfect cosmic ray removal, localized anomalies in the wavelength solution, etc.} we shall demonstrate that our RVs are entirely consistent with the DRS RVs, albeit with better nominal precisions. 

\subsection{\mystar}\label{sec:mystar}

\mystar\ is an $\rm{m_V}=6.36$ G5\;V star with an activity level ($\lrhk\lesssim-5.05$) that would classify it as being inactive or very inactive, depending on the definition \citep[][]{hoffleit87,lazaro97,maldo10}. It has been well studied both photometrically and spectroscopically \citep{arevalo99,chen01,kidger03}. It is a long-term target of the California Planet Search, has no known planets, and has been shown using Keck/HIRES measurements to be stable with rms scatter of $2.6$~\mps\ \citep{jojo09,weiss-thesis16}. It has also been used as a standard star in efficiency and stability testing of HARPS-N \citep{cosentino12}, and has been shown with HARPS-N to be stable at the $\sim2$~\mps\ level, which incidentally is significantly lower than the Sun even after the RV signals due to Solar System planets have been removed \citep{Milbourne19}.

We applied our new RV extraction technique to $735$ HARPS-N spectra of \mystar, taken between 2012 May 23 and 2018 January 19. The median exposure time for these spectra was $300$~s (a little over half of the spectra had $300$~s exposures), with the $\pm\sigma$ range covering exposure times from $120$~s to $450$~s. However, the first $80$-odd spectra were taken over the course of just three nights in May 2012, and almost all of the first $100$ spectra had exposure times of $120$~s. Additionally, the first $130$ spectra were taken with HARPS-N's original CCD, which suffered a broken amplifier on 2012 September 28. The next fifteen spectra were taken during the period of commissioning the replacement CCD, which ended on 2013 February 17, and all the remaining spectra were taken after the new CCD had been commissioned.

\subsection{Comparison of GP and DRS RVs}\label{sec:GP_vs_DRS_RV}

A representative example of a GP fit to a randomly-selected portion of an \mystar\ spectrum is shown in Fig.~\ref{fig:GP_RV_fit}. The RVs extracted using our GP-based technique are presented alongside the RVs from the DRS pipeline in Fig.~\ref{fig:RV_serial}, where all $735$ RVs are shown, and in Fig.~\ref{fig:RV_serial2}, which zooms in on about $90$ RVs from spectra taken over a baseline of about 10 weeks. Because our technique yields RVs only up to an arbitrary additive constant, both ours and the DRS RVs were first mean-subtracted. The rms scatter of the full set of GP RVs was $2.04$~\mps, while the rms scatter of the DRS RVs was $2.10$~\mps. Superficially, at least, the GP and DRS RVs appear very similar, although it is difficult to draw quantitative conclusions based only on these plots, especially given the long observing baselines, uneven time sampling, and large number of observations.

\begin{figure}
	\begin{center}
		\includegraphics[scale=1]{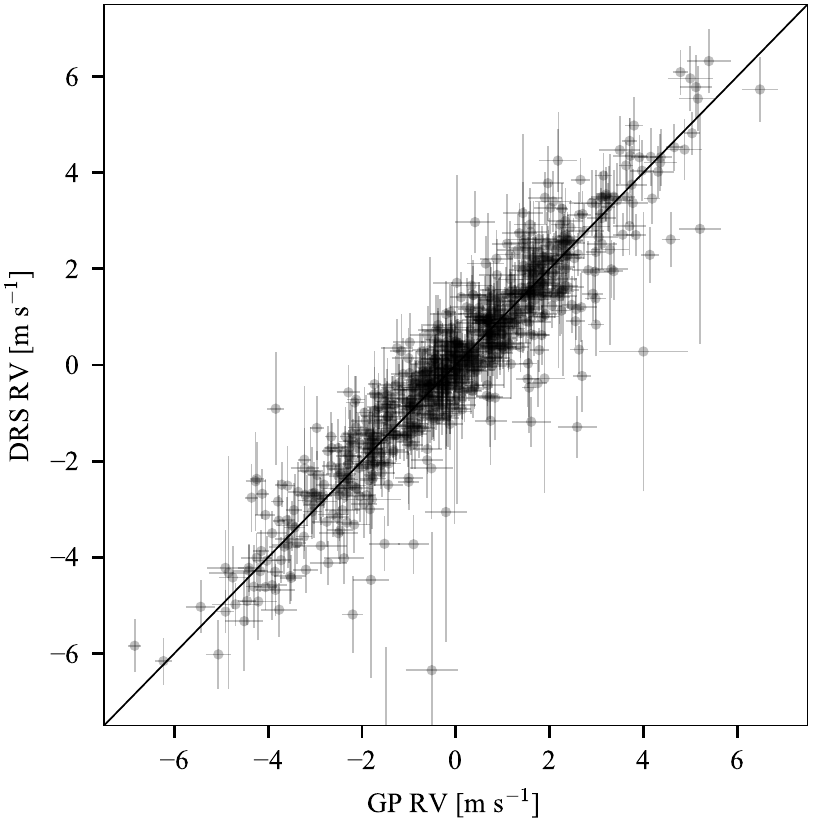}
		\caption[]{RVs extracted using the DRS pipeline (vertical axis) vs.\ RVs from the GP-based technique developed in this paper (horizontal axis). The diagonal line is the 1:1 identity line. The mean \emph{absolute} difference between individual GP and DRS RVs is $57$~\cmps, and $96.5\%$ of the DRS and GP RVs are consistent within $2\sigma$. The mean estimated error for the GP RVs is  $27$~\cmps, vs.\ $60$~\cmps\ for the DRS RVs.}
		\label{fig:RV_vs_RV}
	\end{center}
\end{figure}

\begin{figure}
	\begin{center}
		\includegraphics[scale=1]{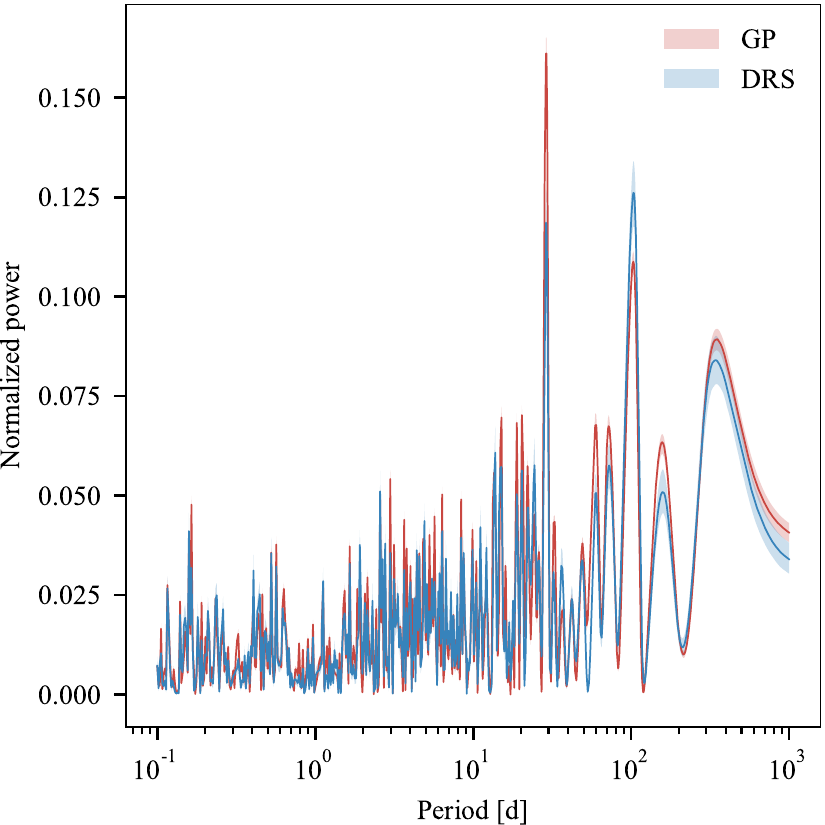}
		\caption[]{Lomb-Scargle power spectra of \mystar\ RVs, as extracted using the GP-based technique developed in this paper (red) and the DRS pipeline (blue). The shaded regions indicate $\pm\sigma$ uncertainty in the power spectra (propagated from the estimated RV errors), which we include to highlight the equivalence of the two power spectra.}
		\label{fig:lomb_scargle}
	\end{center}
\end{figure}

In Fig.~\ref{fig:RV_vs_RV} we plot the DRS RVs vs.\ the GP RVs; we find that the RVs cluster very tightly around the 1:1 identity line, with the mean absolute difference between the individual GP and DRS RVs being only $57$~\cmps\ (median absolute difference $41$~\cmps). In Fig.~\ref{fig:lomb_scargle}, we show the Lomb-Scargle power spectra of both the GP and DRS RVs. The two power spectra are almost indistinguishable; the most prominent peaks in each case are found around $180$~d and $650$~d. There are no periodicities that are present in one set of RVs but not the other. Together, Fig.~\ref{fig:RV_vs_RV} and  Fig.~\ref{fig:lomb_scargle} demonstrate more clearly what the serial representation of the RVs merely suggested, \emph{viz.} that the DRS and GP RVs are remarkably consistent. Given that the DRS RVs were derived using a G2 line mask carefully designed to minimise telluric contamination and maximise RV fidelity, with carefully chosen line weights, while our na\"ive technique fitted everything `blindly,' this is already a favourable first result.

\subsection{Comparison of GP and DRS error estimates}

The mean estimated $1\sigma$ error on the GP RVs was $28$~\cmps\ (standard deviation $8$~\cmps), while the mean estimated error on the DRS RVs was $60$~\cmps\ (standard deviation $31$~\cmps). An important caveat, however, is that our error bars are predicated on the very common though overly-optimistic assumption that the instrument is perfectly stable, or at least that any drift has been perfectly calibrated out.

In Fig.~\ref{fig:cf_mean_error} we plot the mean estimated error in the GP and DRS RVs, as a function of the number of spectra considered ($N=2,\ldots,735$). As anticipated, the error in the GP RVs tends to decrease with increasing $N$, while the error in the DRS RVs remains relatively constant. Of course, despite these broad trends, the mean error in both the DRS and GP RVs does fluctuate, given variable integration times/SNR, atmospheric conditions, etc. Additionally, the GP RV error scaling with $N$ depends on the barycentric velocity correction in a complicated way: more information will be gained by adding those spectra that, after barycentric velocity correction, significantly increases the existing coverage of emission wavelengths (see Fig.~\ref{fig:telluric-shift}). Thus we see that when considering only the first 100 or so spectra, taken over just three nights in May 2012 during which period the wavelength does change very much, the error in the GP RVs hardly improves at all. Elsewhere, much sharper\footnote{The idealised $1/\sqrt{N}$ behaviour assumes, among other things, that all observations have equal errors. Sometimes, improvements that are apparently `better than $1/\sqrt{N}$' are seen in Fig.~\ref{fig:cf_mean_error} (`GP, original order'): this merely reflects the inclusion into the modelling of spectra with a SNR higher than the average SNR of the existing ensemble, which remains encumbered by the low SNR of the first hundred or so spectra.} improvements are seen, while the mean error tends to plateau when including groups of spectra taken in close succession (cf.\ Fig.~\ref{fig:RV_serial}).

To `iron out' the various effects complicating the error behaviour -- in particular, non-random variability in integration times and barycentric velocity shifts -- we also studied the mean error as a function of increasing $N$, but this time where the order of spectra included was randomly permuted. Referring again to Fig.~\ref{fig:cf_mean_error}, we see now that in the case of random permutation, the mean DRS error remains relatively constant no matter how many spectra are included -- since e.g.\ both low and high SNR spectra are added completely at random -- while the GP RV error is seen to decrease very steadily with increasing $N$. The log-log scaling in Fig.~\ref{fig:cf_mean_error} reveals that the GP RV error has a very-nearly monomial dependence on $N$;
%
a fit to the estimated errors suggested $\sigma\propto N^{-0.41}$. 

However, the difference between the GP and DRS error estimates, despite the nearly-identical RV signals, prompts the question: should we believe the GP error estimates? Could they be too optimistic, or the DRS error estimates too pessimistic?

Let us recall that our GP error estimates are computed, essentially, from the observed scatter between RVs derived in many different ways: by comparing different pairs of spectra, and by using different orders within individual spectra. Each RV from one such comparison is associated with its own error estimate, which takes into account photon noise, a model of covariance between nearby points within spectra, interpolation error, the fact that spectral chunks being compared will not in general be identical up to a simple Doppler shift, etc. When we finally impose the constraint that all pairwise RVs thus inferred must in fact be measuring the same thing, and thus average over all of them (refer to equations \ref{eq:avg1}--\ref{eq:avg4}), if it turns out that we are able to home in on the (unknown but well-defined) true average with a precision of e.g.\ $28$~\cmps, then we should be justified in accepting this as a plausible measure of our uncertainty in the inferred RV.

For an $\rm{m_V}=6.36$ G5\;V star and typical integration times of $300$~s, a precision of $\lesssim20$~\cmps\ appears to be more or less the fundamental photon noise-limited precision predicted by \citet{bouchy01}. Moreover, the estimated errors for our GP RVs were larger for shorter integration times ($50$--$60$~\cmps), and smaller for the largest integration times, approaching $12$~\cmps\ for some of the $\gtrsim15$~min exposures. This all suggests that our GP-based error estimates are at least \emph{not} physically implausible.

Further tests will be required to ascertain how our extraction technique compares to the DRS in the case of fainter targets: for example, a precision improvement from $1.8$~\mps\ to $1.5$~\mps\ might seem less impressive than the apparent factor of 2 improvement seen on this data set. Still, even improvements at the level of tens of centimetres per second should be welcomed.

We remark, in passing, that sources familiar with the inner workings of the DRS pipeline inform us that the final DRS error estimates are enlarged by adding approximately $\sim30$~\cmps\ in quadrature to the initial error bars that reflect, essentially, photon noise; the additional error is intended to account e.g.\ for drifts in the wavelength solution and other calibration problems. At face value this might seem to explain the difference between the GP and DRS error bars. However, we note that the pre-quadrature DRS error bars had a mean magnitude of $52$~\cmps; adding $32$~\cmps\ in quadrature would yield a final mean error of $60$~\cmps, as indeed observed. Thus, our GP errors are nearly a factor of 2 smaller even than the pre-quadrature DRS RVs. We also recall that the GP error estimates implicitly contain various contributions from terms not present in the DRS estimates, arising e.g.\ from the imperfect GP fit to the observations, GP interpolation uncertainty, etc. To the best of our knowledge, the DRS errors also do not account for spurious RV signals arising from errors in mask line locations, from varying effective line weights (see Section~\ref{sec:pairwise-xcorr1}), etc.; such issues are not relevant to our technique.

Ultimately, though, the error bars tell us about precision, not accuracy. Both the DRS and our approach may be extracting \emph{some} signal (evidently the same signal) quite precisely: but what \emph{is} that signal? How accurately are we measuring true, dynamical Doppler shifts, as opposed to stellar activity variability, or telluric contamination, or instrumental imperfections? What is the origin of the $\sim2$~\mps\ RV rms scatter shared by the DRS and GP RVs?

\subsection{Some insight into the apparent RV variability}\label{sec:GP_vs_N}

\begin{figure}
	\begin{center}
		\includegraphics[scale=1]{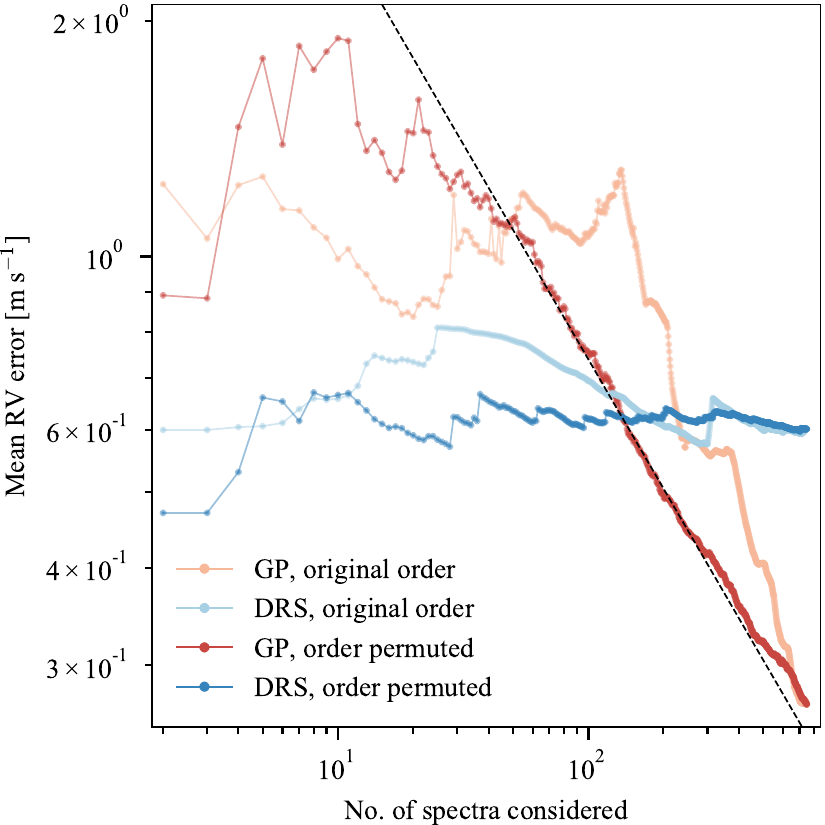}
		\caption[]{Mean (absolute) RV error as a function of $N$, the number of spectra considered. Randomly permuting the order in which spectra are included serves to mask the non-random variability in integration times and observation spacing (the first hundred or so spectra, in chronological order, had very short integration times and were taken over just three nights). For this data set, the GP RV error approaches the ideal $1/\sqrt{N}$ scaling very closely (the dotted diagonal line has slope $-\tfrac{1}{2}$).}
		\label{fig:cf_mean_error}
	\end{center}
\end{figure}

\begin{figure}
	\begin{center}
		\includegraphics[scale=1]{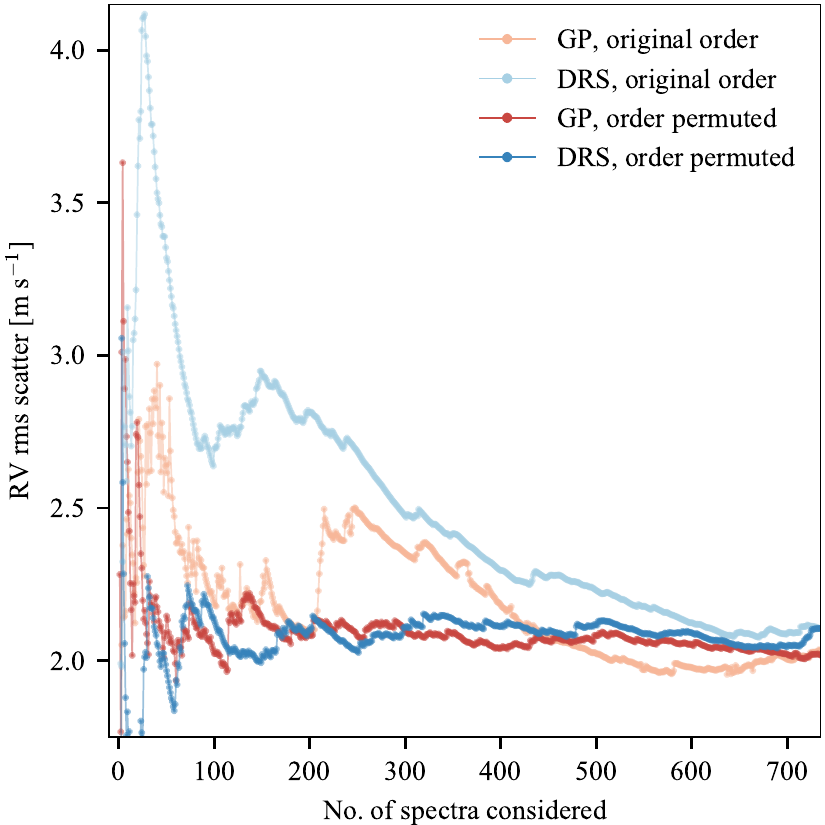}
		\caption[]{As for Fig.~\ref{fig:cf_mean_error}, but for RV rms scatter as a function of $N$. We argue that the finite scatter of $\sim2.1$~\mps\ includes astrophysical and instrumental variability; we should not expect that simply increasing $N$ arbitrarily would lead to a much smaller rms scatter.}
		\label{fig:cf_rms}
	\end{center}
\end{figure}

In Fig.~\ref{fig:cf_rms}, we consider the rms scatter of the GP and DRS RVs vs.\ $N$; as in Fig.~\ref{fig:cf_mean_error}, we show the behaviour both for the case of increasing $N$ in chronological order, and increasing $N$ in a randomly-chosen order. Here we may gain some insight into the very similar final rms scatter of the DRS and GP RVs. We note that when considering only the first 80 or so spectra (in chronological order), the rms scatter in both the GP and DRS RVs is larger than when considering later spectra; the rms scatter peaks at a little over $4$~\mps\ in the DRS case, and just under $3$~\mps\ in the GP case. Recalling that almost all of these initial spectra had exposure times of $120$~s, and were taken over just three nights, we may infer that the increased rms scatter we observe here is partly astrophysical in origin. 

In particular, pressure waves (p-modes) propagate at the surface of Solar-type stars, which leads to a dilation and contraction i.e.\ oscillation of external envelopes over time scales of $5$--$15$~min; the RV signatures of these modes can be as large as several \mps. Granulation phenomena, due to the convective nature of Solar-type stars, can also lead to RV perturbations at the \mps-level, on time scales ranging from minutes to more than a day \citep{dumusque2012}. It is probable that, especially during the initial short-integration observations, we were resolving RV signals due to p-mode oscillations of the G5\;V star \mystar. The subsequent rapid dip in rms scatter may be attributed to a batch of long-integration observations (mostly $900$~s), after which the integration times stabilized around $300$~s. Hence the general decrease in rms scatter observed for large $N$. But since only about half of our observations had integration times long enough to expect efficient ironing out of RV perturbations due to p-mode oscillation, we should expect that the overall RV rms scatter in the GP and DRS RVs would remain finite \citep{Dum11,dum15sun}. Our GP RVs are, so it would seem, simply providing a slightly more precise measure of the finite RV variability of \mystar.

Apart from the short integration times in the earliest observations, we noticed strong indications of large systematic contributions to the rms scatter in the first hundred or so RVs -- possibly though not clearly related to the CCD change after serial observation index $129$. For instance, we noticed several very large jumps ($>4$~\mps) in the ThAr drift correction applied during wavelength calibration for these early spectra: these drifts were about ten times larger than the typical drift corrections in later observations. We also noticed a very strong and statistically significant linear correlation (Pearson coefficient $r\sim60\%$, $p\ll.001$) between these RVs and the applied drift correction, suggesting that the apparent linear trend observed in these early RVs might be spurious. Given the time stamps of the observations, it is possible that some of the variability was related to engineering tests carried out during the initial operation of HARPS-N. Therefore, in the analysis that follows in which we seek to reduce the rms scatter of our GP RVs, we consider only to RVs taken after the CCD replacement, i.e. from observation index $130$ onwards.

Even considering only RVs after the CCD replacement, however, there remained evidence that some of the apparent RV scatter was \emph{not} due to genuine dynamical stellar motion. For example, both the DRS and GP RVs exhibited a strong and significant correlation with the CCF FWHM ($r\sim-38\%$, $p\ll.001$).




\section{Reducing the RV scatter}\label{sec:reducing_rms}

\rev{Now that we have demonstrated in some detail that our technique can yield RVs equivalent to those from the well-regarded DRS pipeline, we turn our attention to a more careful extraction of RVs, and show how simple modifications to our hitherto indiscriminate fitting can significantly reduce the rms scatter of extracted RVs.}

\subsection{\rev{Maximum-likelihood RVs}}

\rev{First, let us eschew working with CCFs in favour of aligning pairs of spectra via direct maximization of likelihood functions. As noted in Section~\ref{sec:real-data}, our sole reason for adopting the former approach was to maximize the consistency between our technique the DRS, i.e.\ to control for any discrepancies in extracted RVs that may have stemmed purely from any issues inherent in  cross-correlation itself. Indeed, we have theoretical grounds for favouring the latter ML technique, and extracting RVs via this approach will also enable a more direct comparison (in Section~\ref{sec:TERRA-SERVAL}) with two other RV extraction codes that rely on maximum-likelihood template matching instead of cross-correlation.}

\rev{We carried out the GP RV extraction as before, but this time replacing equation~\ref{eq:RV-from-CCF} with equation~\ref{eq:RV-from-ML}, to produce ML RV estimates. The rms scatter of these RVs was $2.21$~\mps\ across all RVs, and $1.77$~\mps\ for the RVs taken after the CCD replacement; the corresponding rms scatters were $2.04$ and $1.99$~\mps\ for the GP RVs extracted via cross-correlation. A detailed explanation of why the ML and CCF approaches yield slightly different results, and why the ML approach should be preferred in general, will appear in a brief follow-up paper (Rajpaul~{et al}., \emph{in prep}.). Suffice it to say, however, that in this case at least, the differences between the two sets of RVs were insignificant enough that they could be deemed equivalent to within their formal error bars, in much the same way we argued previously that the DRS RVs were equivalent to our GP RVs extracted via cross-correlation.}

\subsection{\rev{Localizing spectral contamination}}

So far we have done nothing to mitigate any contribution to our RVs from stellar activity or telluric contamination (apart from variance-weighting of RVs from different \'echelle orders): we have fitted all orders in all spectra, with no masking. Yet presumably many if not all orders contain some finite degree of telluric and activity contamination, and perhaps we could better localize such contamination if we sub-divide our orders, and then mask off any sub-orders whose RVs indicate likely contamination, as we did with synthetic data in Section~\ref{sec:synth-data}. 

We investigated splitting each order into $n_\text{split}=2,4,8,\ldots,64$ sub-orders; we preserved the mean functions from the original full order, to avoid introducing spurious RV shifts associated e.g.\ with more weakly-constrained continuum levels. We didn't consider values larger than $n_\text{split}=64$ because otherwise the width of sub-orders would have become comparable to the maximum RV shifts we were fitting ($\pm30$~\kmps, to accommodate shifting of telluric features after barycentric correction), and also because sub-orders would otherwise have been narrower than a few GP covariance length-scales. We then ran our extraction technique as before, but extracted an RV signal for each sub-order, rather than for each full \'echelle order; we computed the statistics suggested in Section~\ref{sec:reduce_contamination}; and we investigated the effect of excluding from the final RV calculation those sub-orders with the unfavourable values of the statistics. 
    
\begin{figure}
	\begin{center}
		\includegraphics[width=\columnwidth]{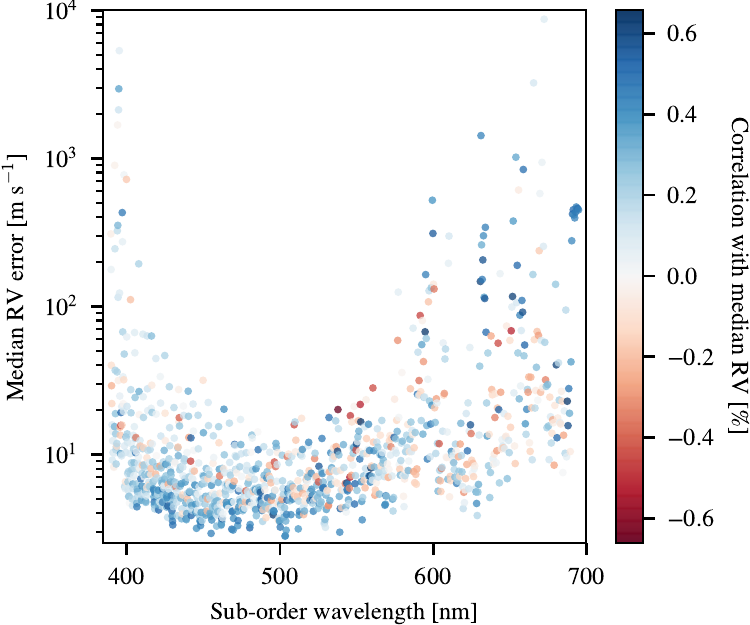}
		\caption[]{Uncertainty in the \mystar\ RV signals extracted from each of a total of \rev{$1\,104$ sub-orders ($n_\text{split}=16$)}; central wavelengths for each sub-order plotted on horizontal axis. The colour scale indicates the linear correlation of the local RVs with the median RV signal across all sub-orders. \rev{See \citet{cretignier2019} for a discussion of how certain lines can give rise to anti-correlated RVs.}}
		\label{fig:order_sig}
	\end{center}
\end{figure}

At first, one might think that maximizing $n_\text{split}$ would lead to best results, since contaminated regions could be resolved as finely as possible in wavelength space. However, by increasing $n_\text{split}$, one decreases the photons available in any sub-order, thus increasing the white noise in the RV signal extracted for that sub-order. Consequently, while very strongly-contaminated sub-orders can indeed be localized more finely, it becomes more difficult to differentiate between RV signals suggesting zero/weak and moderate contamination, since the signal in either case will be dominated by white rather than correlated noise. \rev{Using very large values of $n_\text{split}$ also leads to more significant computer memory burdens. We found that for \mystar, there appeared to be little benefit in using values larger than $n_\text{split}=16$, so we focus hereafter on this case}. 

In Fig.~\ref{fig:order_sig} we plot the median uncertainty in the sub-order RV signals for the case \rev{$n_\text{split}=16$}, and also the correlation of sub-order RV signals with the median RV signal across all orders. Evidently, extracting RVs on such a quasi line-by-line basis yields a wealth of information that would otherwise be lost when studying only a global CCF. For example, we see that especially in the bluest and reddest parts of the spectrum, there are a few sub-orders with RV uncertainties an order of magnitude or more greater than the typical uncertainty. Some of this we are able to identify as likely chromospheric activity (around $395$~nm) or telluric contamination (redwards of $600$~nm). It is possible that poor wavelength solutions also contribute, especially in the red, due to a lack of suitable Th lines and powerful Ar lines bleeding between orders. Especially in the blue, we noticed a spike in RV uncertainty and rms scatter at the edges of many consecutive orders (unfortunately this effect is not visible with the scaling in Fig.~\ref{fig:order_sig}), which may be due to poor wavelength solutions associated with lower ThAr emission flux at order edges. Also apparent in Fig.~\ref{fig:order_sig} is that the majority of the sub-orders with small RV uncertainties are positively correlated with the median RV signal. 

\rev{In a simple approach where we iteratively excluded sub-orders based on their RV rms scatters, we were able to reduce the rms scatter in our ML RVs from $1.77$~\mps\ to $1.66$~\mps\ simply by excluding about half of the sub-orders with the largest rms scatters. Taking the same approach but excluding sub-orders on the basis of their mean RV uncertainties, we were instead able to reduce the rms scatter to $1.69$~\mps. While clearly somewhat effective, this approach assumes that the sub-orders with smallest uncertainties or variability all encode the same signal, which in general will not be true. Importantly, it also assumes that the sub-order RVs have adequate SNR to permit identification of contamination; in general, low SNR might make it necessary to consider instead \emph{combinations} of sub-orders.}

\rev{In a more flexible though still heuristic approach, we started with a small set of sub-orders with small uncertainties, then selected a random sub-order for inclusion in that set. We then checked whether the rms in the RVs extracted from that set of sub-orders decreased; if not, we selected a different random sub-order for inclusion until the rms decreased, at which point we repeated the process by including further sub-orders, until no further improvements appeared possible.}

\rev{Across multiple runs of this algorithm, we were reliably able to reduce to RV rms scatter to below $1.30$~\mps. The best result we obtained, out of twenty runs, was a final rms scatter of $1.21$~\mps, obtained when using roughly a third of all available sub-orders (in other words we `threw away' two thirds of the sub-orders). The corresponding mean uncertainty in these RVs was $57$~\cmps.

Note that this result certainly does \emph{not} represent the truly optimal combination of sub-orders: after all, the total number of ways to choose an arbitrary subset from a set of $L$ sub-orders is $\sum_{l=1}^{L} {L \choose l}$, which for $L=1\,104$ means $\sim10^{332}$ possible combinations; we explored tens of thousands of combinations, at best. Nevertheless, the $1.21$~\mps\ scatter in our `decontaminated' RVs represents a $\sim30\%$ improvement on our initial RVs, while maintaining a precision on par with the DRS RVs.}

While we regard this reduction in RV scatter as extremely encouraging -- \rev{and, following the arguments we made in Section~\ref{sec:reduce_contamination}, we can take this as indicative of us having removed red/correlated noise from the RVs} -- further work is needed to understand the exact nature of the apparent contamination we are removing. It would be interesting, for example, to compare the sub-orders we excluded with the line lists from e.g.\ \citet{dumusque18}. Given the wealth of information available when extracting RV signals for each sub-order, there are many other ways we could gain insight into spectral contamination. For example, we could try to study local broadening or asymmetries in RV likelihood functions. We could also consider chromatic variations in RVs, e.g.\ by studying the difference in RVs from a subset of blue and red orders. Since genuine Doppler signals due to stellar reflex motion should be encoded identically in all orders, these signals should disappear when differencing red and blue RVs, whereas instrumental, telluric and activity variability (the latter due, among other factors, to spot contrast being strongly wavelength dependent) would not.

\subsection{\rev{Comparison with HARPS-TERRA and SERVAL}}\label{sec:TERRA-SERVAL}

\rev{We now make a few quantitative comparisons between our decontaminated GP RVs, the DRS RVs we considered throughout Section~\ref{sec:real-data}, and RVs we extracted using the publicly-available HARPS-TERRA \citep{anglada12} and SERVAL \citep{zech2018} codes.\footnote{As the DRS RVs may be considered equivalent to our `pre-decontamination' RVs, we do not include the latter RVs in these comparisons.} Both of the latter codes derive RVs by least-squares alignment (essentially an ML approach) of observed spectra with a template derived from the same observations; for more information, see also the discussion in Section~\ref{sec:comparison}. 

Table~\ref{tab:RV_summary} summarizes the rms variation and mean estimated error in the RVs extracted using the four different methods. We note that our GP RVs exhibit the smallest variation, with estimated errors on par with the other methods.

\begin{table}
	\caption{\rev{Summary of rms variation and mean estimated error in the \mystar\ RVs extracted using four different methods.}}
	\label{tab:RV_summary}
	\begin{tabular}{lcc}
		\hline
		& RV rms & Mean estimated error\\
		& [\mps] & [\mps]\\
		\hline
		DRS & 1.95 & 0.55 \\
		GP (decont.) & 1.21 & 0.57 \\
		HARPS-TERRA & 1.71 & 0.58\\
		SERVAL & 1.82 & 0.71 \\
		\hline
	\end{tabular}
\end{table}

\begin{table}
	\caption{\rev{Summary of the overlap, to within $1\sigma$ and $2\sigma$ error bars, between the \mystar\ RVs from the four different extraction methods in Table~\ref{tab:RV_summary}.}}
	\label{tab:RV_overlap}
	
	\begin{tabular}{lcccc}
		\hline
		 (1$\sigma$ agreement)& DRS & GP (decont.) & HARPS-TERRA \\
		 \hline
		 GP (decont.) & 52\% & - & -  \\
		 HARPS-TERRA & 59\% & 67\% & -  \\
		 SERVAL & 82\% & 61\% & 71\%  \\
		\hline
	\end{tabular}

	\begin{tabular}{lcccc}
	\hline
	(2$\sigma$ agreement)& DRS & GP (decont.) & HARPS-TERRA \\
	\hline
	GP (decont.) & 75\% & - & -  \\
	HARPS-TERRA & 89\% & 95\% & -  \\
	SERVAL & 99\% & 87\% & 97\%  \\
	\hline
	\end{tabular}


\end{table}

Fig.~\ref{fig:RV_vs_RVs} plots the DRS, HARPS-TERRA and SERVAL RVs vs.\ our GP RVs. As in Fig.~\ref{fig:RV_vs_RV}, we find that the RVs cluster reassuringly around the 1:1 identity line, with the GP RVs being most closely aligned with the HARPS-TERRA RVs. This is further quantified in Table~\ref{tab:RV_overlap}, which shows that $67\%$ of the GP and HARPS-TERRA RVs are consistent within their respective $1\sigma$ error bars (and $95\%$ within $2\sigma$). The SERVAL and DRS RVs show an even higher degree of consistency, with about $82\%$ overlapping within $1\sigma$. This degree of overlap would be unlikely to arise if the different extraction methods were producing profoundly different RV signals, and/or if the error bars on the RVs were significantly underestimated.

\begin{figure}
	\begin{center}
		\includegraphics[scale=1]{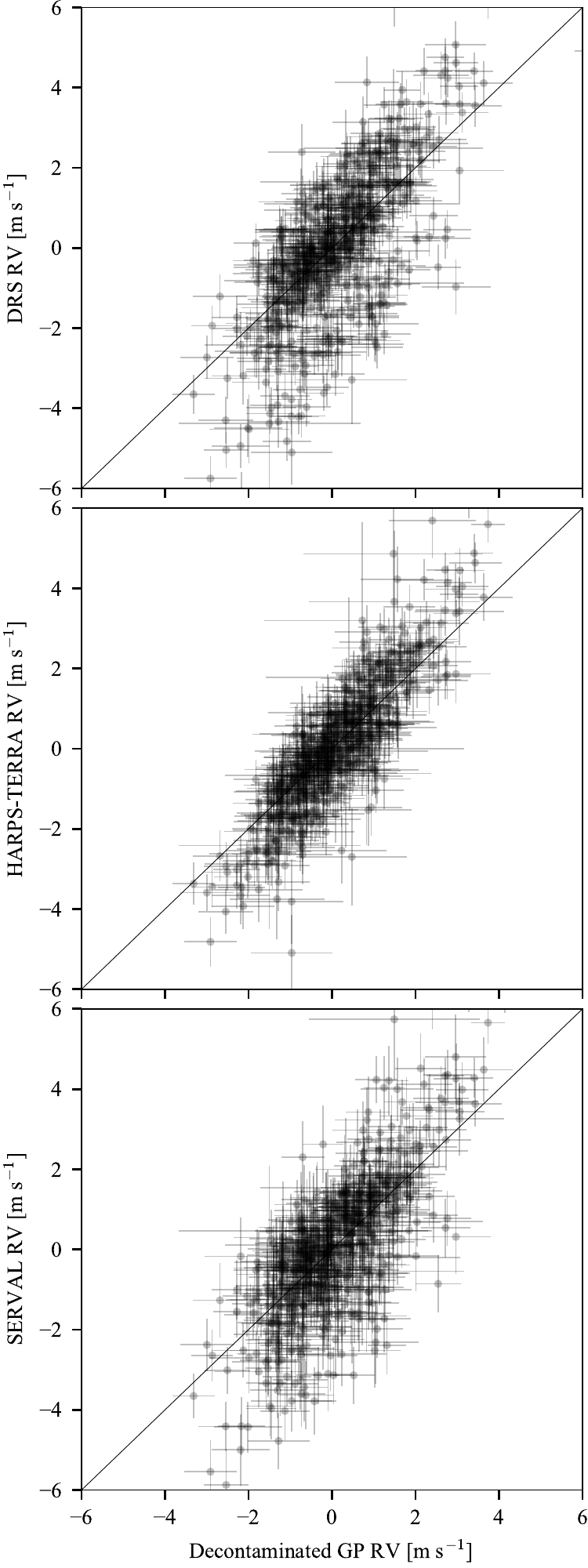}
		\caption[]{\rev{As for Fig.~\ref{fig:RV_vs_RV}, but now showing (from top to bottom) the DRS, HARPS-TERRA, and SERVAL \mystar\ RVs, all vs.\ our decontaminated GP RVs. The mean (median) absolute difference between our GP RVs is $1.03$~$(0.75)$~\mps\ for the DRS; $68$~$(54)$~\cmps\ for HARPS-TERRA; and $94$~$(72)$~\cmps\ for SERVAL.}}
		\label{fig:RV_vs_RVs}
	\end{center}
\end{figure}


Yet there are, evidently, differences between the four sets of RVs. This is not surprising, given the divergent approaches they take to RV extraction. How then do we make sense, physically, of these differences? Table~\ref{tab:RV_correlations}, which presents linear correlation coefficients between the RVs and various time-varying quantities, sheds some light on the matter.

\begin{table*}
	\caption{\rev{Linear correlation between \mystar\ RVs extracted using four different methods (rows) and various time-varying quantities (columns) that are expected to be independent of stellar RVs, though possibly sensitive to stellar activity or telluric contamination. For each possible correlation, we give both the Pearson coefficient, $r$, and the $p$-value. Correlations that are significantly nonzero at the $0.05$ level are shown in bold.}}
	\label{tab:RV_correlations}
	\begin{tabular}{lcccccccccccc}
		\hline
		 - & \multicolumn{2}{c}{CCF FWHM} & \multicolumn{2}{c}{CCF BIS} & \multicolumn{2}{c}{\lrhk} & \multicolumn{2}{c}{BERV} & \multicolumn{2}{c}{Air mass} & \multicolumn{2}{c}{Exposure time}   \\
		 &$r$&$p$ &$r$&$p$ &$r$&$p$ &$r$&$p$ &$r$&$p$ &$r$&$p$ \\
		 \hline

 		 DRS &$\boldsymbol{-37.6\%}$&$\boldsymbol{\ll.001}$ &$-0.7\%$&$.87$ &$-2.9\%$&$.48$ &$\boldsymbol{-13.4\%}$&$\boldsymbol{.001}$ &$\boldsymbol{-8.9\%}$&$\boldsymbol{.03}$ &$+4.3\%$&$.30$  \\
 		 
 		 GP (decont.)&$\boldsymbol{+12.4\%}$&$\boldsymbol{.002}$ &$+1.1\%$&$.79$ &$-0.2\%$&$.97$ &$-4.6\%$&$.26$ &$+0.9\%$&$.83$ &$+5.0\%$&$.23$  \\
 		 
 		 HARPS-TERRA&$\boldsymbol{-8.4\%}$&$\boldsymbol{.04}$ &$+0.2\%$&$.95$ &$+0.4\%$&$.93$ &$\boldsymbol{-13.2\%}$&$\boldsymbol{.001}$ &$-7.6\%$&$.07$ &$+7.3\%$&$.08$  \\
 		 
 		 SERVAL &$\boldsymbol{-31.7\%}$&$\boldsymbol{\ll.001}$ &$-5.8\%$&$.16$ &$+1.5\%$&$.71$ &$+7.4\%$&$.06$ &$-5.5\%$&$.18$ &$+5.8\%$&$.16$  \\
		\hline
	\end{tabular}
\end{table*}

We see that both the DRS and SERVAL RVs appear to exhibit strong and statistically-significant linear correlations with the CCF FWHM; this correlation is weaker and absent in the GP and HARPS-TERRA RVs, respectively. No significant correlations are found between any of the RVs and the CCF BIS or the \lrhk\ index, although the DRS and HARPS-TERRA RVs appear weakly correlated with the air mass at the time of observation, and all RVs except for the GP ones appear to exhibit moderate correlations with the Earth's barycentric velocity at the time of observation.\footnote{\rev{The Lomb-Scargle power spectra of the DRS, HARPS-TERRA, and SERVAL RVs contained significant power around $180$~d, with the HARPS-TERRA RVs also containing significant power around $360$~d, suggesting annual variability in these RVs.}} Taken in tandem, this suggests the presence of a lower degree of activity and telluric contamination in the GP RVs than in those extracted by the other methods.

Of course, a linear correlation coefficient measures only one very simple type of relationship between variables, and a negligible linear correlation does not imply the absence of a more general relationship. Conversely, linear correlation tests are known to be sensitive to outliers \citep{bin1990}, although this seems unlikely to be a major issue given the small overall amplitude of RV variability. Indeed, when we replaced Pearson linear correlation with a more robust (Spearman) rank correlation test, no new correlations appeared between the GP RVs and the various other time series, but additional and significant correlations did appear between the SERVAL RVs and the CCF BIS, and between the DRS RVs and the \lrhk\ index. In short, these correlation tests strongly \emph{suggest} though do not prove that our GP RVs contain less spectral contamination than the other RVs. Further work is needed to understand in detail the origin of the differences between RVs from the different codes.
}

\section{Discussion}\label{sec:discuss}

\subsection{Comments on practical aspects of our implementation}

The results in the preceding sections were obtained using a set of Python scripts we wrote to process a group of HARPS-N spectra and ultimately yield a set of RVs and error estimates. In total, the main scripts -- including pre-processing steps such as applying barycentric velocity and instrumental drift corrections; \rev{GP-based pairwise spectral alignment}; error propagation; etc.\ -- spanned some hundreds of lines of code. We expect our implementation could be made faster though various optimizations, e.g.\ by porting the slowest parts of our scripts from native Python to faster  \textsc{Fortran} or C++ equivalents, or by leveraging massively-parallel GPU computation to speed up pairwise spectral alignment. Some forethought could also make it more compact, especially if one were to make use of existing libraries for tasks such as GP regression (we wrote our own GP regression code for this project, if only to put us in a better position to diagnose any problems we found with the implementation). The results we have presented were derived using what was essentially a crude and inelegant but at least carefully-tested initial implementation. We intend in future to make publicly available a stand-alone version of our main code, provisionally named \textsc{grace}, to stand for Gaussian process RV Activity Extraction.

In any case, we note that deriving RVs \emph{ab initio} from $735$ \mystar\ spectra took about two days on a modern desktop with $16$~GB RAM and a $4.2$~GHz quad-core CPU. The greatest computational expense was associated with comparing a little over half a million different pairs of spectra; each such comparison involved manipulating GP posterior distributions evaluated at hundreds of thousands of wavelengths.

Two days is, evidently, orders of magnitude longer than one might reasonably expect from a more conventional RV extraction approaches. However, in practice -- even if using an implementation as crude and inefficient as our initial one almost certainly was -- this should not be a significant obstacle. First, most target stars will not be observed nearly as many times as the one considered here. For a star observed 100 times, say, there would be a factor of fifty times fewer pairwise comparisons to be made, and so one could expect to extract the RVs from scratch in about an hour, rather than a couple of days. Second, whenever a new observation is added to a set of existing ones, it is not necessary to repeat all pairwise comparisons: it is merely necessary to compare all existing spectra with the new one, and to repeat the averaging calculations. If one were to add a new observation to the 735 \mystar\ spectra analyzed here, computing an RV for the new spectrum (and updating all the old RVs) would take a matter of minutes.

\subsection{Conceptual comparison with existing techniques}\label{sec:comparison}


\rev{
The defining features of our approach to RV extraction are (i) pairwise RV extraction, without construction or referencing of of any global template; (ii) the use of GPs to model and align spectra; and (iii) selective aggregation of localized (quasi line-by-line) RV estimates, in order to remove spectral contamination. As we discuss below, each of these features has appeared, albeit possibly in isolation, in some guise in other RV extraction techniques. If nothing else, our technique does at least appear to be unique in unifying these advantageous features into a conceptually simple framework that seems to work extremely well on both synthetic and real data sets.}

For example, the idea of enforcing some sort of self-consistency when comparing multiple pairs of spectra to derive RVs is not new; ideas regarding combining cross-correlations first appeared in \citet{zucker03} and \citet{zucker06}. While \citet{zucker06} give a useful derivation of the $1/\sqrt{N}$ limiting behaviour we have discussed, they hit up against an RV precision limit after co-adding 10--15 spectra in the tests they present, and do not seem able to outperform standard techniques. This might be partly due to issues inherent in cross-correlation, to their lack of a scheme for performing principled interpolation that accounts for the non-trivial covariance structure of spectra, or to their lack of a strategy for mitigating spectral contamination. The work of \citet{prieto07} similarly foreshadowed aspects of our technique in terms of performing multiple pairwise comparisons, but their technique seeks only to improve RVs that have already been measured by some other technique.


\rev{Our technique is also hardly the first to deliberately avoid using a synthetic template or template from a different star, and to favour instead using only the science observations themselves to construct a template; this approach has been taken elsewhere for both} absorption cell spectrographs \citep{sato02,johnson06,gao16} and stabilized spectrographs \citep{anglada12,defru15,zech2018}, \rev{and has been demonstrated using both CCFs and ML (or least squares) approaches. HARPS-TERRA and SERVAL, which we used for RV extraction in Section~\ref{sec:TERRA-SERVAL}, both adopt such a least-squares template matching approach.} Compared to our technique, most of these schemes generally contain parameters that need to be hand-tuned, make simplifications for the sake of theoretical or numerical tractability, and/or entail \emph{ad hoc} interpolation schemes e.g.\ for co-adding observations at different wavelengths. More fundamentally, however, they are hindered by a `chicken and egg' problem: ascertaining the spectral regions exhibiting most variability would require knowing the true stellar RV, so that all spectra could first be shifted to a common velocity scale; but doing so would require already knowing the true stellar RV, without any activity contribution (see discussion in Section~\ref{sec:pairwise-xcorr1}). \rev{We circumvent these issues by avoiding completely the construction of a global template, whether once-off or iteratively,} and instead use pairwise comparison to derive an ensemble of differential RVs.


\citet{czekala17} presented a technique using GPs to disentangle spectroscopic binaries; they noted that the technique could also be profitably applied in the context of exoplanet detection. Their technique shares with ours all of the advantages offered by GP regression, but in the sense that they do not rely on pairwise spectral alignment, their technique is \emph{much} more computationally intensive. Some of the results they present are based only on narrow bandpasses selected from about ten high-resolution spectra -- even then, the computation takes about a day on a computer cluster. This is comparable to the time it took us to derive RVs from $735$ spectra, using the full available wavelength range, on a desktop computer. Such difficulties are not insurmountable, but our technique has the advantage of theoretical simplicity and computational efficiency.


\rev{A few other studies have assessed the relative weight that might be given to different \'echelle orders to empirically account for stellar variability and telluric contamination, including though not limited to the HARPS-TERRA paper by \citet{anglada12}. Recently,}  \citet{dumusque18} individually fitted the positions of thousands of individual lines in each spectrum, resulting in a quasi-independent, low-precision RV time-series for each line fitted. The individual lines were chosen from a binary masks used in a `standard' RV extraction process, and RVs for each line were computed by comparison with a template derived by co-adding all spectra. RVs from all the individual lines was then combined by computing the inverse variance weighted average after $\sigma$-clipping to remove outliers. While this approach showed similar results and performance to the DRS, it is considerably less data-driven than our own approach -- specifically, it still relies on the line-list encoded in the cross-correlation mask, which is not guaranteed to be optimal for any given target. Furthermore, as we have already emphasized, building a template by simply co-adding observed spectra is sub-optimal. Nevertheless, the approach does enable a study of the dependence of the RVs derived from individual lines on e.g. stellar activity. \cite{dumusque18} showed, for example, that the rms scatter of \rev{2010 HARPS RVs of Alpha Cen B (47 nights of observation spanning 81 days)} can be reduced from $1.95$~\mps\ to $1.21$~\mps\ by using only 25\% of the line list, specifically those least correlated with the RVs derived from the full set of lines.

\rev{In a follow-up study extending the work of \citet{dumusque18}, \citet{cretignier2019} showed that Alpha Cen B spectral lines of different depths in the stellar atmosphere are affected differently by stellar activity, and estimated that at least $89\%$ of the individual lines are linearly correlated with the overall activity signal, which they assumed to be activity-dominated. By choosing a subset of lines uncorrelated with the putative activity signal, or by using the difference in RVs from lines formed at different depths as an activity proxy, they were able to reduce the RV rms scatter to $\sim0.8$--$0.9$~\mps. These results are extremely encouraging; it would be interesting to see how our method fares at reducing the rms scatter in a single observing season of RVs from a relatively active star such as Alpha Cen B (recall that \mystar, by way of comparison, is an inactive star, and we achieved a $1.21$~\mps\ rms scatter on RVs spanning several years)}.

\rev{Finally, it is worth comparing our GP-based method to a particularly sophisticated approach to dealing with one type of spectral contamination: \emph{viz}., that taken by} {\sc wobble}  \citep{wobble19}, which focusses on deriving precise RVs in the presence of telluric absorption. {\sc wobble} models all spectra simultaneously as a combination of a stellar spectrum, which does not change over time but whose RV does, and one or more telluric components, which scale with air mass but remain fixed in the observatory's rest frame. The observed spectra (or rather their logarithms) are modelled as a linear combination of these components, within a ML framework. The model has many thousands of parameters: the individual stellar and telluric components in each pixel, the RV shifts of the star,  the relative contribution of the telluric components at each epoch, and regularisation constants, which are used to avoid overfitting the stellar and telluric spectra. \citet{wobble19} demonstrate the performance of their method on several real datasets, and their results are extremely similar to the DRS RVs. The code is highly optimized and uses TensorFlow to implement the core linear algebra calculations very efficiently, considering the complexity of the model. However, the reliance on regularisation is one of the main drawbacks of this approach, both conceptually (since the physical interpretation of the effect of the regularisation is unclear) and computationally (\citeauthor{wobble19} report that tuning the regularisation parameters is the slowest step). {\sc Wobble} shares a number of important characteristics with the GP RV method presented in this paper: its data-driven philosophy, the absence of templates for either the stellar or the telluric spectra, and the fact that the model produces high-SNR, over-sampled templates as a bonus. Like our GP method, {\sc wobble} models all observations simultaneously and the RV precision (as well as the quality of the extracted templates) \emph{should} improve as the number of observations increases, although \citet{wobble19} did not investigate this aspect in detail. Unlike our GP method, {\sc wobble} does not make use of the covariance between neighbouring pixels in the spectra. The main conceptual advantage of {\sc wobble} over other RV extraction methods, including ours, is of course the fact that it explicitly models time-variable telluric absorption. Furthermore, \citet{wobble19} mention that it might be possible to extend the model to include time-dependent behaviour in the stellar spectrum associated with stellar activity. 


\subsection{Future work}\label{sec:future}

The work we have presented here lends itself naturally to a number of extensions.

First, in this paper we used a crude scheme for identifying regions of spectra that are likely contaminated by activity or telluric variability. While this scheme worked surprisingly well, one should be able to do better. \rev{On a simple level, one might measure localized broadening of or asymmetries in likelihood functions (or CCFs), during pairwise spectral comparison, to probe stellar activity contamination. Or one might use a mixture model to decompose localized (sub-order) RVs into putative stellar reflex, stellar activity, telluric, and possibly other components. A more sophisticated approach would be to extend our GP covariance kernel to include a temporal dependence in addition to a wavelength dependence.} Thus one could model explicitly and simultaneously any time-varying changes in observed spectra due e.g.\ to stellar activity, and to disentangle these effects from much simpler RV shifts. 
The chief obstacle in such an approach to generalizing our GP model would be computational cost: covariances would need to be formulated between \emph{all} available spectra, rather than just between observations within single spectra or pairs of spectra. Initial tests with a small number of spectra provided us with proof of this concept, i.e., we found that it \emph{does} appear possible to use a GP formulation to disentangle time-varying changes in spectral shape from simple RV shifts, and thus to avoid entirely the need to mask off any portions of spectra. Nevertheless it is clear that direct, joint GP modelling of many hundreds of \'echelle spectra will almost certainly be computationally intractable, even when leveraging sparse matrix inversion algorithms. We are currently investigating computationally efficient strategies for constraining multiple time-varying components (RV shifts, activity, tellurics) of a large set of observed spectra. 

Second -- in a much simpler extension than the aforementioned one -- we could enforce temporal covariance structure on the RVs themselves, rather than the spectra from which they are derived. For example, we should not expect a star's RV to change by several~\kmps\ over the course of an hour, though a change of a few \mps\ might well be reasonable; in short, observations closely spaced in time should be associated with similar RVs. Imposing such covariance structure should allow us to derive slightly more accurate and precise RVs compared to the case where we ignore the time at which an observation was made, and pretend that the RV we assign to a given spectrum is completely independent of the RVs we assign to other spectra closely-spaced in time.


Third, this paper has focused on precise RV extraction; we did not actually study the  stellar template we could construct from the GP models of the individual \mystar\ spectra. Such a low-noise, high-resolution stellar template could be of interest in its own right e.g.\ for better constraining the astrophysical properties of a given target. Ideally, the template spectrum should be constructed not via simple co-adding, but rather by evaluating the GP posterior distribution over \emph{all} spectra simultaneously. A small caveat is that the effect of inter-/intra-pixel sensitivity variations on the super-resolved spectrum may need to be considered.   

Fourth, we have presented results here of RV extraction from spectra of an extremely unremarkable Solar analogue star, for which excellent templates already exist (as evidenced, among others, by the precision of the DRS RVs). As our technique is completely agnostic to stellar type, it would be worth applying it to targets for which canonical templates do not already exist, e.g.\ young T Tauri stars, brown dwarfs, and emission line stars \citep{blake2010,rei2018,gaia_dr2_2018}. It is also well known that the delta function template technique does not perform well with M-stars, which typically have many molecular lines in their spectra \citep{anglada12,defru15}. We also intend to apply our technique, in the near future, to targets where different spectrographs lead to apparently inconsistent RVs -- most notably Kepler-10 \citep{Rajpaul2017} -- with a view to gaining insight into the origin of such inconsistencies.

Lastly, we have used here a single large data set to compare the real-world performance of our technique with that of an industry-standard extraction techniques, and with theoretically-predicted precision limits. Our results were encouraging, \rev{but more work is needed to understand the relative extent to which our favourable results are due to pairwise RV extraction, GP modelling, and/or careful recombination of localized RV estimates; such understanding might well be used to simplify our technique even further. It would also} be worth comparing our technique with the performance of a number of different RV extraction techniques on well-studied stars (e.g.\ Alpha Cen B), since each technique may be expected to have different strengths and weaknesses. Given that for every star except one we are ignorant of the true RV signal, some sort of blind RV extraction challenge using realistic synthetic data might be a useful long-term goal to pursue within the RV community. Alternatively, Solar spectra could provide an ideal data set to use for comparing different RV extraction techniques \citep{dum15sun}, although even with the Sun it seems it might be very difficult to reduce apparent RV variability below the \mps~level \citep{Milbourne19}.

\section{Conclusions}\label{sec:conc}

We have presented in this proof-of-concept paper a simple new approach for extracting differential stellar RVs, which seeks to address many of the drawbacks associated with  \rev{more conventional approaches to RV extraction}. In short, we model each stabilized spectrum using a Gaussian process (GP), align each GP model with every other GP model spectrum, and use simple algebraic operations to aggregate the pairwise RVs thus inferred. Our technique enforces the requirement that while we may align a given spectrum with any number of others serving as templates, we must ultimately assign a single, well-defined RV to the spectrum in question. In the limit of many observations, our method thus yields RVs equivalent to the case where a perfect template existed. Errors associated with template mismatch or noisy, finite-resolution observed templates are eliminated; our technique doesn't depend on any pre-defined line lists or mask; it provides a direct and meaningful way to quantify errors in extracted RVs; and, by working with a large ensemble of spectra and extracting RVs on a highly localized basis, it \rev{suggests ways} to mitigate activity and telluric contamination. Moreover it could be extended to allow telluric and stellar activity contamination to be more modelled in more sophisticated ways. Above all, though, it is both simple and transparent: the basic technique can be implemented in a few hundred lines of code, it requires no hand-tuning for different instruments or stellar types, and doesn't even require the input of any astrophysical knowledge.


We used tests with synthetic data to demonstrate that our method works as intended. We then derived RVs from $735$ HARPS-N spectra of the inactive  G5\;V star, and compared our RVs with RVs derived by the HARPS-N DRS pipeline \rev{(based on template cross-correlation), as well as the HARPS-TERRA and SERVAL codes (based on least-squares matching between observations and a template derived from the same observations). We showed that an initial, `blind' RV extraction with our code yielded RVs equivalent to, though with better nominal precisions, than the DRS RVs. We then showed that by excluding certain spectral regions (likely associated with stellar activity or telluric variability) from our RV calculations, we were able to reduce the RV rms scatter by about $30\%$, or over half a \mps, resulting in a precision on par with but an rms scatter smaller than the RVs derived by any of the other codes.}


We noted that our technique does lend itself to a number of extensions, the most interesting and powerful of which would be explicit joint modelling of stellar activity variability alongside simpler RV shifts. Nevertheless, apart from providing a method of deriving RVs that serves as a useful and independent check on RVs from standard (possibly far more complex and opaque) pipelines, we have shown that our technique even in its existing form can derive RVs with \rev{favourable precisions and apparent contamination levels compared to other widely-used approaches}. Thus our technique seems well placed to enable the study of smaller planets around a wider variety of stars than is currently possible. 

Particularly decisive benefit may be derived in the case of targets observed very many times. For example, the Terra Hunting Experiment on HARPS3 will aim to observe carefully selected targets almost daily for ten years, with a view to detecting Earth-twins with RV semi-amplitudes around $10$~\cmps\ \citep{HARPS3,hall2018}; our technique should attain its best possible performance in this case of very large $N$ and long observing baselines. Given the ground-breaking planet detection goals of a programme such as the Terra Hunting Experiment, it will be imperative to optimize not only instrumental performance and observing strategies, but also the extraction techniques used to deliver the RVs through which Earth-twins will, hopefully, be discovered.

\section*{Acknowledgements}

\rev{The authors wish to thank the referee, Guillem Anglada-Escud\'e, for many valuable comments that helped us improve our manuscript.} The authors are grateful to the HARPS-N science team for making available to us the \mystar\ spectra studied in this work. VMR acknowledges the Royal Astronomical Society and Emmanuel College for supporting this work. SA acknowledges support from STFC via grant ST/N000919/1.




\bibliographystyle{mnras}
\bibliography{biblio}



\appendix

\section{Covariance kernels}\label{sec:app-kernel}
If we start from the reasonable assumption that a given flux's value could be related to the values of fluxes at nearby wavelengths, but that any such informativeness should grow weaker as the distance between fluxes grows, we arrive at a family of so-called `stationary' covariance functions that depend only on $|\lambda-\lambda'|$, i.e.\ the Euclidean distance (in wavelength, or more generally in time, space, etc.) between observations in a spectrum. 

By far the most widely-used example of a stationary kernel is the \rev{Gaussian or} squared exponential (SE) covariance kernel:
\begin{equation}
{k_{{\rm{SE}}}}({\lambda },{\lambda' }) = {h^2}\exp \left( {\frac{{ - {{({\lambda } - {\lambda' })}^2}}}{{2{\rho ^2}}}} \right).
\end{equation}
The SE kernel is governed by two so-called `hyperparameters,' $h$ and $\rho$, which respectively govern the output (e.g.\ flux) and input (e.g.\ wavelength) scales of functions. Note that the term \emph{hyperparameters} is used as these parameters govern the properties of a GP prior distribution over functions, rather than acting explicitly on the functional form of curves drawn from the distribution. Functions drawn from an SE kernel will be infinitely differentiable, and will exhibit smooth variations with a typical length scale for variations of $\rho$. 

Despite the SE kernel's very widespread use, \citet{stein1999} notes that the very strong smoothness assumptions underpinning the SE kernel are often unrealistic for modelling physical processes in the real world, for which the Mat\'ern class of may instead be more appropriate. The Mat\'ern class of covariance functions is defined by
\begin{equation}\label{eq:matern-kernel}
{k_\nu }({\lambda },{\lambda'}) = {h^2}\frac{{{2^{1 - \nu }}}}{{\Gamma (\nu )}}{\left( {\frac{{\sqrt {2\nu } |{\lambda } - {\lambda '}|}}{\rho }} \right)^\nu }{\mathbb{B}_\nu }\left( {\frac{{\sqrt {2\nu } |{\lambda} - {\lambda '}|}}{\rho }} \right),
\end{equation}
where $h$ and $\rho$ are the output and input scales, as before; $\Gamma()$ is the standard gamma function extending the factorial function; and $\mathbb{B}()$ is the modified Bessel function of second order. The hyperparameter $\nu$ controls the degree of differentiability of the resultant functions modelled by a GP with a Mat\'ern covariance kerel, such that they are only $\nu+\tfrac{1}{2}$ times differentiable. 

As $\nu\to\infty$,  the functions become infinitely differentiable, and the Mat\'ern kernel becomes the SE kernel.  \citet{rasmussen2006} note that in the absence of explicit prior knowledge about the existence of higher order derivatives, in practice it is probably not feasible to use a finite set of noisy data to distinguish between finite values of $\nu\ge \tfrac{7}{2}$ and $\nu\to\infty$. At the other end of the spectrum, setting $\nu=\tfrac{1}{2}$ gives us the Ornstein-Uhlenbeck (OU) or exponential kernel: this kernel looks almost identical to the SE kernel, but the ${{{({\lambda } - {\lambda'})}^2}}$ term in the exponential is replaced with $\left| {{\lambda } - {\lambda '}} \right|$, and the functions it models are \emph{not} smooth. Instead, they are only once differentiable, and correspond to the random motion of massive Brownian particles under the influence of friction. The Mat\'ern kernel with $\nu=\tfrac{5}{2}$, advocated in Section~\ref{sec:kernel}, describes functions which are -- roughly speaking -- somewhat smooth, but which still permit fairly sharp changes, as observed e.g.\ in spectral absorption or emission lines.

While we show in this paper that $\nu=\tfrac{5}{2}$ Mat\'ern kernel does a surprisingly good job of fitting stellar spectra, it is worth noting that function draws from such a kernel (somewhat-smooth, correlated random noise), when not conditioned on stellar spectroscopic data, will usually not bear very close resemblance to a typical stellar spectrum (absorption lines imprinted on a continuum, say). It seems possible, then, that there may exist better ways for leveraging GPs to model stellar spectra, either by finding more appropriate kernels, or by transforming observed stellar fluxes in some way so that spectra more closely resemble functions generated by a standard kernel. 


\bsp	
\label{lastpage}
\end{document}